\documentclass[aps,prb,twocolumn,showpacs,floatfix,superscriptaddress,10pt]{revtex4-1}
\usepackage{graphicx, color}
\usepackage{tabularx}
\usepackage{overpic}
\usepackage{amsmath, amsthm, amssymb}
\usepackage{empheq}
\usepackage{bm}
\usepackage{multirow}
\usepackage{hyperref}

\newcommand{\mb}[1]{\mathbf{#1}}
\newcommand{\pd}{\partial}

\newcommand{\mbb}[1]{\mathbb{#1}}
\newcommand{\mc}[1]{\mathcal{#1}}
\newcommand{\pmat}[1]{\begin{pmatrix}#1\end{pmatrix}}

\def\sgn{\textrm{sgn}}
\newcommand{\comments}[1]{} 
\begin{document}
\title{Distant-Neighbor Hopping in Graphene and Haldane Models}
\author{Doru Sticlet}
\email{sticlet@pks.mpg.de}
\affiliation{Laboratoire de Physique des Solides, CNRS UMR-8502, Univ.\ Paris Sud, 91405 Orsay, France}
\affiliation{Max-Planck-Institut f\"ur Physik komplexer Systeme, D-01187 Dresden, Germany}
\affiliation{LOMA, CNRS UMR-5798, Univ.\ Bordeaux, F-33400 Talence, France}
\author{Fr\'ed\'eric Pi\'echon}
\affiliation{Laboratoire de Physique des Solides, CNRS UMR-8502, Univ.\ Paris Sud, 91405 Orsay, France}
\begin{abstract}
Large Chern number phases in a Haldane model become possible if there is a multiplication of Dirac points 
in the underlying graphene model. This is realized by considering long-distance hopping integrals. 
Through variation of these integrals, it is possible to arrive at supermerging band touchings, 
which up to N7 graphene are unique in parameter space. They result from the synchronized motion of all supplementary Dirac points into the regular $\pm\mb K$ points of graphene. 
The energy dispersion power law is usually larger than the topological charge associated with them. 
Moreover, adding distant-neighbor hoppings in the Haldane mass allows one to sweep large Chern number phases in the topological insulator.
\end{abstract}
\maketitle

\section{Introduction}
The Haldane model\cite{Haldane} is the first topological insulator that presents the quantum Hall effect without 
an external magnetic field. It is a two-band system with bands that has a nontrivial topology. The bands are 
characterized by Chern numbers that are proportional to the conductance carried by edge states. This model 
was a playground for ideas that eventually led to the prediction and discovery of the $\mathbb Z_2$ topological insulators.\cite{KaneQSH, Bernevig}

Here we revisit the Haldane model and show in practice how the addition of hopping integrals between 
distant neighbors can lead to a multiplication of topological phases with a large Chern number. 
Recent work suggests a way to produce flat bands with arbitrary Chern numbers in multilayer 
systems.\cite{Trescher, Yang} In contrast, we constrain ourselves to the two-band Haldane system 
and we do not seek flatness of bands. Admittedly this is not a very physical way to increase 
the topological index characterizing a band, because the contribution from distant neighbors 
are small. The conceptual advantage is that one can fully describe the phase diagram 
of such systems and analytically predict its topological transitions. 
The system can be understood from a decomposition of the model in an underlying gapless graphene model and a Haldane mass.
 The variation of the topological index is related at once to the multiplication of nodes in the energy dispersion 
for the gapless system and to the rapid oscillations in the Haldane mass term.
 In general, if the two-band underlying gapless system admits $2n$ Dirac points, 
then the Chern number can vary from $-n$ to $n$.\cite{Sticlet} 

In Sec.~\ref{sec:two} we present the theoretic tool to compute analytically 
the Chern number in a two-band system. That allows one to immediately discriminate 
the topological phases. In Sec.~\ref{sec:three}, we treat the underlying graphene 
with long-distance hopping integrals. Up to N7 (next $\times$ 6-nearest-neighbor) graphene, 
we investigate the multiplication of Dirac points through the addition of long-range hopping. 
We also show that there are unique supermerging points where all the Dirac points merge
 to $\pm\mb K$ points in the graphene Brillouin zone (BZ).
Satellite Dirac points can be found by perturbing around these special band touchings. 
The topological charge associated with them can be immediately established 
from a sum over the additional Dirac points. 
In Sec.~\ref{sec:three}, we consider the effects of gapping the Dirac points with a Haldane mass term. 
The phase diagram for the modified Haldane model is shown.

\section{Chern number in two-bands models}\label{sec:two}

A generic two-band Hamiltonian on a two-dimensional Bravais lattice reads
\begin{equation}
H=\frac{1}{4\pi^2}\int_{\rm BZ}d^2\mb k \! \sum_{\alpha,\beta=1,2} h_{\alpha \beta}(\mb k)c^{\dag} _{\alpha \mb k} c_{\beta \mb k}
\end{equation}
with $c^{\dag} _{\alpha \mb k}$ the creation operator of the Bloch state with wave vector $\mb k$ and where $\alpha$ constitutes a pseudospin index 
resulting from either two sublattices or two orbitals per unit cell.
The elements $h_{\alpha \beta}(\mb k)$ form a $2\times 2$ Hermitian matrix $h(\mb k)$ that can be written
\begin{equation}
h(\mb k)= \sum_{\mu=0}^3 h_\mu(\mb k)\sigma_\mu,
\end{equation}
with $\sigma_0$ the identity matrix and $\sigma_{1,2,3}$ the Pauli matrices.
 $h_{\mu=0,3}(\mb k)$ comes from {\em intrasublattice} contributions $h_{\alpha \alpha}$ and 
 $h_{\mu=1,2}(\mb k)$ from {\em intersublattice} contributions $h_{\alpha \beta}$.
The real valued functions $h_{\mu}(\mb k)$ can be further split into even and odd components $h_{\mu}(\mb k)=h_{\mu}^e(\mb k)+h_{\mu}^o(\mb k)$, where $h_{\mu}^e(\mb k)=h_{\mu}^e(-\mb k)$
and $h_{\mu}^o(\mb k)=-h_{\mu}^o(-\mb k)$. For time-reversal symmetric, spinless Hamiltonians, $h_{\mu=0,1,3}(\mb k)$ are purely even and $h_{2}(\mb k)$ purely odd.

The spectral decomposition of matrix $h(\mb k)$ reads
\begin{equation}
h(\mb k)=\sum_{\pm } \epsilon_{\pm}(\mb k)P_{\pm}(\mb k),
\end{equation}
with band energies $\epsilon_{\pm}(\mb k)=h_0(\mb k) \pm |\mb h(\mb k)|$ and eigenband projector 
$P_{\pm}(\mb k)=\frac{1}{2}(\sigma_0\pm \mb h\cdot\bm\sigma/|\mb h|)$,
where $\mb h(\mb k)=(h_1,h_2,h_3)$. Component $h_0(\mb k)$ breaks particle-hole symmetry by shifting the energy bands and it may also lead to an indirect overlap of the two energy bands. Nevertheless it does not intervene in the direct gap $|\mb h|$ or in the projectors $P_{\pm}$ which determine the topological properties of the Hamiltonian.
In the following we will neglect the $h_0(\mb k)\sigma_0$ term and consider the system an insulator as long as the direct gap $|\mb h|$ does not close; in this situation the projection to the lower band is always well-defined.
An insulating phase, in which the three components of $\mb h(\mb k)$ never vanish simultaneously 
and $|\mb h|$ remains finite for any $\mb k$, can be characterized by a topological index, the first Chern number $\mc C$. The integer $\mc C$ counts how many times the Brillouin zone wraps around the unit sphere traced by $\hat{\mb h}=\mb h/|\mb h|$. 
One can choose to index the system with the Chern number associated with the lowest band $\epsilon_{-}(\mb k)$,
\begin{equation}\label{intch}
\mc C=\frac{1}{4\pi}\int_{\rm BZ}d^2\mb k  \ \hat{\mb h}
\cdot(\pd_{k_x}\hat{\mb h}\times\pd_{k_y}\hat{\mb h}),
\end{equation}
where the integral is over the Brillouin zone.
The Chern number is zero unless one of the component of $\mb h(\mb k)$ breaks time-reversal symmetry.
Furthermore, a non-zero value of $\mc C$ requires that any submodel obtained by considering only two components of $\mb h$ 
must exhibits band touchings at some finite set of points in the BZ.\cite{Sticlet}
In fact, an explicit calculation of $\mc C$ is made easy by considering such a gapless system containing 
only two components of $\mb h$. When the band touchings of the gapless submodel have linear 
dispersion (i.e. they are Dirac points), $\mc C$ can be calculated by treating separately the chirality 
of the Dirac points and the sign the third component of $\mb h$ (the {\em mass term} that gaps the system) at these Dirac points.
The Chern number then reads\cite{Sticlet}
\begin{equation}\label{chno}
{\mc C}=\frac{1}{2}\sum_{\mb k\in D_i} \chi_i(\mb k)\,\sgn(h_i),
\end{equation}
where $D_i$ is the set of Dirac points for a simplified two-component model without an $h_i$ term,
and
\begin{equation}\label{chirality}
\forall\mb k\in D_i,\quad\chi_i(\mb k)=\sgn\big(\pd_{k_x}\mb h\times\pd_{k_y}\mb h\big)_i
\end{equation}
is their corresponding chirality.
Such formula permits an analysis of the Haldane model by separately 
studying the underlying gapless graphene model and the sign of the mass term. 
The caveat of the above formula is that it works only for point band touchings with linear energy dispersion (Dirac points). 
However, we shall see that
there can be Fermi lines or point band touchings with higher power dispersion as well. The latter ones can be understood as the merging of many Dirac points. 
Then the topological charge of the merging points is just the sum of the chiralities for the Dirac points that are converging to it. 
This fast calculation of charge associated with a band touching will be referred to as the sum rule.\cite{Volovik1}

\section{Distant-neighbor hopping in graphene}

\begin{figure}[t]
\centering
\includegraphics[width=\columnwidth]{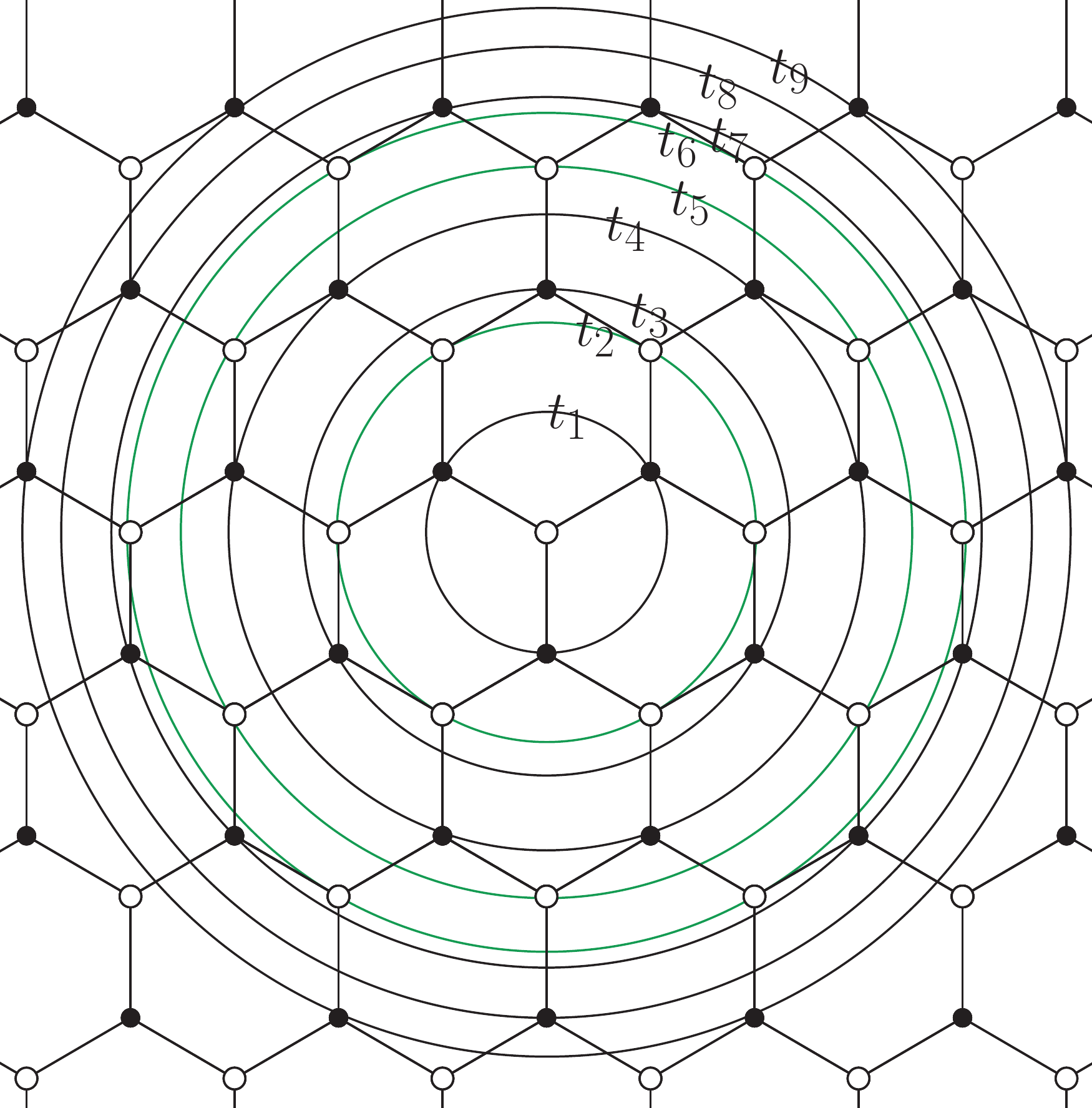}
\caption{The possible hoppings in the graphene N9 model. From a central B atom, the neighbors are arranged in concentric circles. The hopping integrals from the central B atom to a site placed on a circle is denoted by $t_n$, $n$ growing with the distance between sites.}
\label{fig:hexhop}
\end{figure}

As seen in previous section, the possibility of a two-band insulator with a Chern number  $\mc C=n$ 
requires one to build a reduced two-band gapless model with at least $2n$ Dirac points. Let us consider the extended tight-binding graphene model,
including distant N$n$ [(next $\times(n-1)$-nearest-neighbor]
hopping terms on the hexagonal lattice.
The eventual Dirac point will eventually be gapped by a Haldane mass to yield a topological insulator with large Chern numbers.
The hexagonal lattice is a bipartite lattice built out of two interpenetrating triangular Bravais sublattices $A$ and $B$. Let us denote by $t_n$ the (isotropic) N$n$ intra- and intersublattice hopping terms (see Fig.~\ref{fig:hexhop}).
The parameter $t_1$ corresponds to usual nearest-neighbor N1 graphene.
In this section we considered only intersublattice hoppings $t_n$ in units of $t_1$, such that there are $n-1$ free parameters.
The intrasublattice hopping terms contribute to the identity Pauli matrix $\sigma_0$ and are neglected in the following. 
For real-valued $t_n$, the matrix $h(\mb k)$ 
preserves time-reversal and inversion symmetries [e.g. $h^*(-\mb k)=h(\mb k)$ and $\sigma_1 h(-\mb k) \sigma_1= h(\mb k)$)]. 
Moreover, when considering only intersublattice hopping $t_n$, there is a sublattice symmetry characterizing the system. The symmetry is represented by the operator $\sigma_3$ which anticommutes with $h(\mb k)$.\cite{Ryu2002}
Explicitly, $h(\mb k)$ reads
\begin{equation}
h(\mb k)=\pmat{0 & f(\mb k)\\ f^*(\mb k) & 0},
\end{equation}
with $f(\mb k)=h_1(\mb k)-ih_2(\mb k)$ or
\begin{equation}
f(\mb k)=\sum_n t_n g_n(\mb k),
\end{equation}
where the functions $g_n(\mb k) $ up to N9 are tabulated in Table~\ref{tab:tab2} in which
$\mb a_1=\sqrt{3}a\big(\frac{1}{2},\frac{\sqrt{3}}{2}\big)$ and $\mb a_2=\sqrt{3}a\big(-\frac{1}{2},\frac{\sqrt{3}}{2}\big)$
denote the primitive vectors of the triangular sublattices. 
The hexagonal lattice constant $a$ is set to 1 from now on.
The explicit form of the function $g_n(\mb k)$ corresponds to a Bloch basis such that $g_n(\mb k+\mb G)=g_n(\mb k)$ 
with $\mb G$ a reciprocal lattice vector.

\begin{table*}
\caption{Properties of N$n$ $AB$ intersublattice hopping terms. 
Physical distance is counted in units of lattice constant $a$.
Chemical distance is the smallest number of bonds passed while hopping between two sites. In the ``neighbors'' column are the number of sites counted at a given physical distance from a chosen central site. In contrast, note that any site has $3n$ neighbors located at a {\em chemical} distance $n$ from it. The primitive vectors of the triangular sublattice are
$\mb a_1=\sqrt{3}a\big(\frac{1}{2},\frac{\sqrt{3}}{2}\big)$ and $\mb a_2=\sqrt{3}a\big(-\frac{1}{2},\frac{\sqrt{3}}{2}\big)$.  
}
\begin{center}
\footnotesize
\begin{tabular}{c c c c c c}
\hline\hline
\multirow{2}{*}{N$n$}& \multirow{2}{*}{Hopping} & Physical & Chemical & \multirow{2}{*}{Neighbors} & \multirow{2}{*}{$g_n(\mb k)$} \\
& & distance & distance & & \\
\hline
N1&$t_1$ & 1 & 1 & 3 & 
$1+e^{-i\mb k\cdot\mb a_1}+e^{-i\mb k\cdot\mb a_2}$  \\
N3&$t_3$ & 2 & 3 & 3 & 
$e^{i\mb k\cdot(\mb a_1-\mb a_2)}+e^{i\mb k\cdot(\mb a_2-\mb a_1)}+e^{-i\mb k\cdot(\mb a_1+\mb a_2)}$\\
N4&$t_4$ & $\sqrt{7}$ & 3 & 6 & 
$e^{i\mb k\cdot\mb a_1}+e^{i\mb k\cdot\mb a_2}
+e^{-2i\mb k\cdot\mb a_1}+e^{-2i\mb k\cdot\mb a_2}
+e^{i\mb k\cdot(\mb a_1-2\mb a_2)}+e^{i\mb k\cdot(\mb a_2-2\mb a_1)}$ \\
N7&$t_7$ & $\sqrt{13}$ & 5 & 6 &
$e^{i\mb k \cdot (2\mb a_1-\mb a_2)} +e^{i\mb k \cdot (2\mb a_2-\mb a_1)}
+e^{2i\mb k \cdot (\mb a_1-\mb a_2)} +e^{2i\mb k \cdot (\mb a_2-\mb a_1)}
+e^{-i\mb k \cdot (2\mb a_1+\mb a_2)} +e^{-i\mb k \cdot (2\mb a_2+\mb a_1)}$\\
N8&$t_8$ & 4 &5 & 3 &
$e^{i\mb k \cdot (\mb a_1+\mb a_2)}+e^{i\mb k \cdot (\mb a_1-3\mb a_2)} +e^{i\mb k \cdot (\mb a_2-3\mb a_1)}$\\
N9&$t_9$& $\sqrt{19}$ &5 & 6 &
$e^{-3i\mb k \cdot \mb a_1} +e^{-3i\mb k \cdot \mb a_2}
+e^{i\mb k \cdot (2\mb a_1-3\mb a_2)} +e^{i\mb k \cdot (2\mb a_2-3\mb a_1)}
+e^{2i\mb k \cdot \mb a_1} +e^{2i\mb k \cdot \mb a_2}$\\
\hline\hline
\end{tabular}
\normalsize
\end{center}
\label{tab:tab2}
\end{table*}

\subsection{Nearest-Neighbor N1 graphene review}

\begin{figure}[t]
\begin{overpic}[width=\columnwidth]{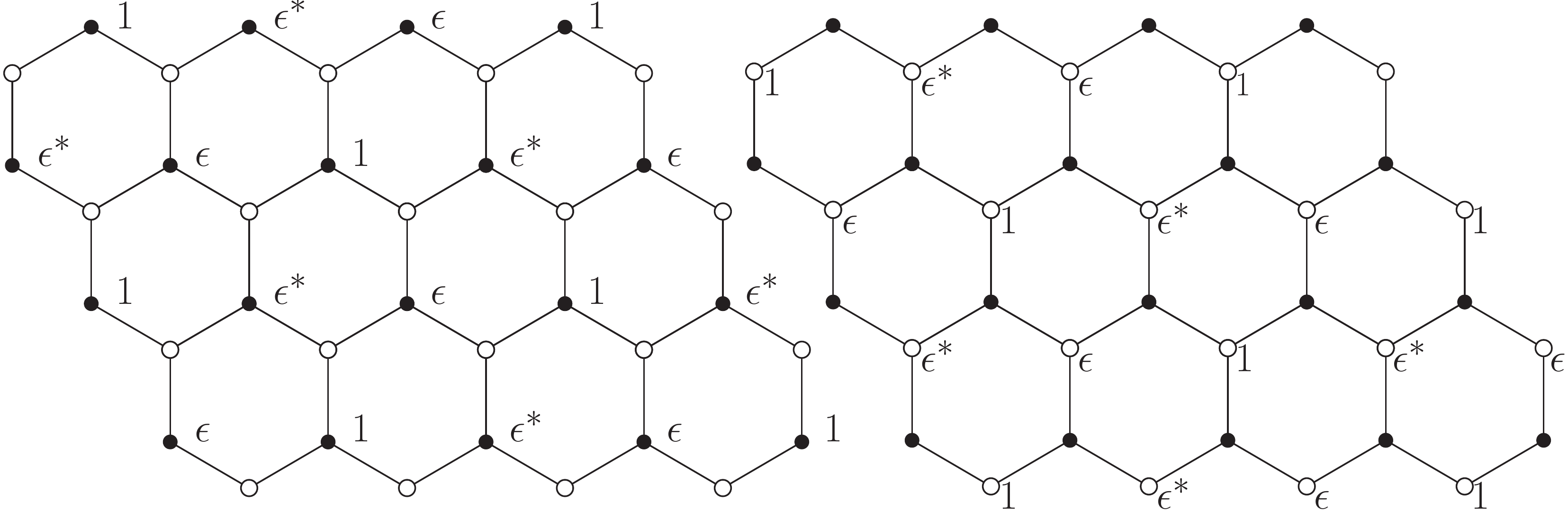}
\put(0,30){(a)}
\put(47,30){(b)}
\end{overpic}
\caption{Real space representation of the four zero-energy eigenstates of N1 graphene.\cite{Lukyanchuk2008} Filled (empty) bullets represent A (B) sublattice sites.
Wave functions components on different lattice positions are related by Bloch theorem $\psi_{\mb k}(\mb r+\mb R)=e^{i\mb k \cdot \mb R}\psi_{\mb k}(\mb r)$ with $\mb R$ 
any Bravais lattice vector. Let us denote $\epsilon=e^{i\mb K\cdot\mb a_1}=e^{2\pi i/3}$; then  $e^{i \mb K\cdot\mb a_2}=\epsilon^*$ 
with $1+\epsilon+\epsilon^*=0$ and $\epsilon^3={\epsilon^*}^3=1$. 
Left figure corresponds to valley $\mb K$ and A sublattice and right figure to valley 
$\mb K$ and B sublattice.
The wave functions in $-\mb K$ valley are obtained by complex conjugating the amplitudes at $\mb K$.
Wave function amplitudes are invariant under $C_3$ rotation around a lattice site and under translations ($\mb R_{\perp}=m(\mb a_1+\mb a_2)$) perpendicular to $\mb K$ and exhibit periodicity under translations parallel to $\mb K$ with a period $\mb R_{\parallel}=3(\mb a_1-\mb a_2)$.
Time reversal exchanges valley index without changing sublattice index. Inversion exchanges both valley and sublattice indices. 
The product of the two operations thus exchange sublattice index only. 
}
\label{fig:N1zerostates}
\end{figure}

Before studying N$n$ graphene, let us briefly review the usual properties of N1 graphene where $f(\mb k)=t_1 g_1(\mb k)$.
The two energy eigenvalues are given by $\epsilon_{\pm}(\mb k)=\pm |f(\mb k)|$
and there is a gap separating the two bands.
Band touch\-ings occur at isolated positions $\pm\mb K$ corresponding to zeros of $f(\mb k)$.
For N1 graphene the zeros correspond to the two nonequivalent Brillouin zone corners $\pm \mb K=\pm (\mb a_1 ^*-\mb a_2 ^*)/3$
where $\mb a_1^*=\frac{4\pi}{3a}\big(\frac{\sqrt{3}}{2},\frac{1}{2}\big)$ and $\mb a_2^*=\frac{4\pi}{3a}\big(-\frac{\sqrt{3}}{2},\frac{1}{2}\big)$.
At each of these points, there are two degenerate zero-energy eigenstates. As illustrated in Fig.~\ref{fig:N1zerostates}, the bipartite property allows one to project one state entirely on the $A$ sublattice and the other on the $B$ sublattice.\cite{Lukyanchuk2008}
Altogether there are four zero-energy states, each labeled by two indices:
a valley index corresponding to $\pm\mb K$ and a sublattice index $A$ or $B$ equivalent to the eigenvalues $\pm$ associated with sublattice symmetry operator $\sigma_3$. 
The time-reversal transformation exchanges valley index without changing sublattice index.
Inversion (represented by $\sigma_1$) exchanges both valley and sublattice indices.
Hence the product of the two operations exchanges sublattice index only.

In the neighborhood of the band touchings $\pm\mb K$, one can expand the function $f(\mb k)$ in small momenta $\mb q=q(\cos{\theta},\sin{\theta})$. It follows that $f(\pm \mb K+\mb q)\simeq qe^{\mp i\theta}$ and 
the linearity in $q$ identifies the band touchings as massless Dirac fermions. More generally, if the band touching has $f\propto (q e^{-i\theta})^n$, then its respective chirality is $n$. This translates to the fact that the two dimensional vector $(h_1,h_2)\propto q^n(\cos(n\theta),\sin(n\theta))$ rotates counterclockwise by $2\pi n$ for $\theta$ sweeping once the interval $[0,2\pi)$. Here, for $n=1$, it follows immediately that Eq.~(\ref{chirality}) and the low-energy expansion both predict $\chi(\pm\mb K)=\pm 1$.

\subsection{Band touchings in \texorpdfstring{N$n$}{Nn} graphene}

Let us study how the properties of the zero-energy states are modified when distant-neighbor hoppings are considered.
For N$n$ graphene, band touchings occur at positions $\pm\mb k$ corresponding to zeros of $f(\mb k)=\sum_n t_n g_n(\mb k)$.
Previous solutions, $\pm\mb K$, obey $g_n(\pm \mb K)=0$ and thus remain zeros of $f(\mb k)$ regardless of the new hopping integrals $t_{n>1}$.
To find the positions of other zeros, one can keep $\mb k$ on the three high-symmetry lines $T$ joining $\Gamma$, $\pm\mb K$, and $M$.
These lines are globally invariant under time reversal, $C_3$, $C_2$ and inversion with respect to the $\Gamma$ point.
Without loss of generality, let us analyze the $T$ line, $\mb k=k(1,0)$ (see Fig.~\ref{fig:move}). Along it $g_n(\mb k)$ 
is a real function and the condition $f(\mb k)=0$ translates into a polynomial equation 
$h_1(x)=0$ for the variable $x=\cos(\frac{\sqrt{3}}{2}k)$.
Up to N8 this polynomial reads as
\begin{eqnarray}\label{points}
h_1(x)&=&4\big(x+\frac{1}{2}\big)p(x),\quad\text{with}\notag\\
p(x)&=&4(t_7+t_8)x^3+ 2(t_4-t_8)x^2+(t_3-4t_7-3t_8)x\notag\\
&&+\frac{1}{2}-\frac{t_3}{2}-t_4+t_7+\frac{3t_8}{2}
\end{eqnarray}
The maximum number of solutions is given by the degree of the polynomial and it clearly increases with the range of hopping, 
but not systematically since N7 and N8 correspond to a polynomial with the same degree. When all the solutions are distinct, they correspond to band touchings with linear dispersion in the $k_x$ direction.
As anticipated $x=-\frac{1}{2}$ (at $\pm \mb K$) is a solution regardless of the value of $t_n$. 
Other physically meaningful solutions must verify $|x|\le 1$. 
For each such solution upon applying a $C_3$ rotation, one can associate three band-touching points at 
$\mb k_1=k(1,0)$, $\mb k_2=k(-\frac{1}{2},\frac{\sqrt{3}}{2})$ and $\mb k_3=k(-\frac{1}{2},-\frac{\sqrt{3}}{2})$. 
From time-reversal symmetry, it follows that there are three additional touching points at $-\mb k_{1,2,3}$.
Hence a solution $|x| \le 1$ with $x \ne \{\pm 1,-1/2 \}$ implies at least  six nonequivalent band-touching points at $\pm \mb k_{1,2,3}$. 
In contrast, a solution $x=-1$ is associated with a single touching point at $\Gamma$, 
a solution $x=-\frac{1}{2}$ to the two nonequivalent BZ corner $\pm\mb K$ 
and a solution $x=1$ to the three nonequivalent $M$ points.
In a N$n$ graphene model with a polynomial $h_1(x)$ of degree $m\le n$ the maximum number of nonequivalent band-touchings points 
per valley on a $T$ line is thus $[1+ 3(m-1)]$.
For example, in N7 graphene the degree of the polynomial is $m=4$, and therefore there are a maximum of ten Dirac points per valley.
As a final remark concerning the band-touching points, we stress that we cannot exclude the possibility of having additional touching points outside the high-symmetry 
$T$ lines. However, the only case encountered in the numerical simulations is that of Fermi lines (zero energy lines) 
which connect the zero-energy solutions located on the $T$ lines. 
These are particular solutions that can be expected when a nondegenerate zero on 
the $T$ line exhibits a vanishing chirality (see Sec.~\ref{subsec:N4graphene} for an example). 

\subsection{Zero-energy state wave functions in \texorpdfstring{N$n$}{Nn} graphene}

\begin{figure}[t]
\includegraphics[width=\columnwidth]{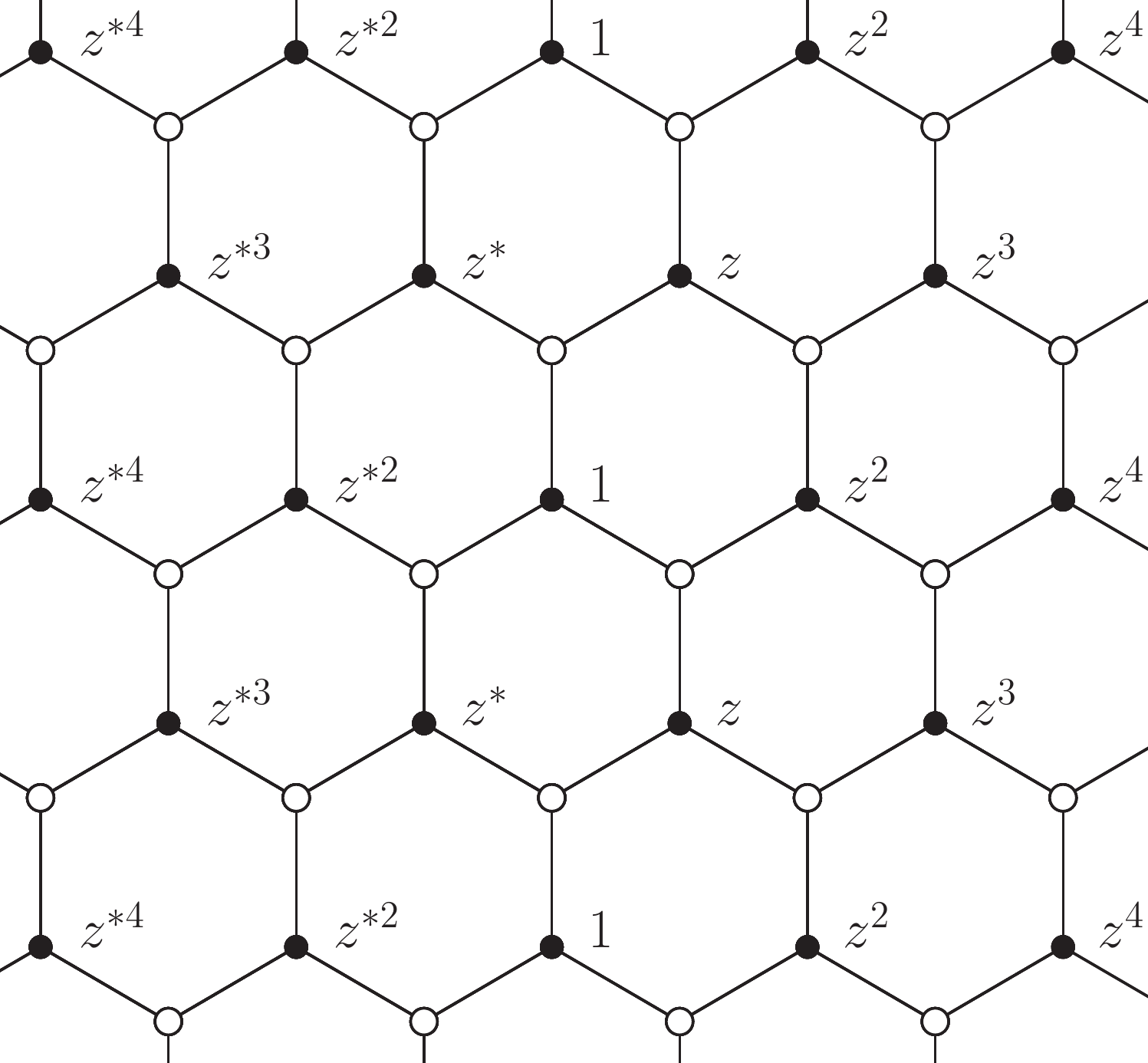}
\caption{Real-space representation of a generic zero-energy eigenstate of N$n$ graphene projected on the sublattice $A$ (in valley $\mb K$). 
Wave functionscomponents on different lattice positions are related 
by Bloch theorem $\psi_{\mb k}(\mb r+ \mb R)=e^{i\mb k \cdot \mb R}\psi_{\mb k}(\mb r)$ with $\mb R$ 
any Bravais lattice vector. For $\mb k$ on high-symmetry lines we have $z=e^{i \mb k \cdot \mb a_1}$ and $z^*=e^{i \mb k \cdot \mb a_2}$
(for $\mb k\ne\pm\mb K$, $1+z+z^* \ne 0$ and $z^3 \ne 1\ne{z^*}^3$).
The three additional states are obtained by performing a $C_3$ rotation  
around the center of a hexagon. Solutions in the opposite valley are obtained  by complex conjugating the amplitudes.
 A similar picture can be drawn for states projected of the $B$ sublattice.
}
\label{fig:Nnzerostates}
\end{figure}

Similarly to N1 graphene, there are two degenerate zero-energy eigenstates that correspond to each band-touching point of N$n$ graphene.
The bipartite property is still valid for N$n$ graphene and it allows one
to project one zero-energy state on the $A$ sublattice and the other on the $B$ sublattice.
The real-space representation of these two energy states is illustrated in Fig.~\ref{fig:Nnzerostates} for a generic $\mb k=k(1,0)$ on a $T$ line.
The wave function exhibits translation invariance in the direction perpendicular to $\mb k$ and it is multiplied by a phase 
$z^2=e^{2i \mb k \cdot \mb a_1}$ on both $A$ and $B$ sublattices when translated by one unit along $\mb k$. Figure~\ref{fig:Nnzerostates} 
is especially useful as it allows one to quickly construct the polynomials $h_1(x)$ at all orders.

\subsection{Velocities and chirality of Dirac points in \texorpdfstring{N$n$}{Nn} graphene}

When all the band-touching points on a given $T$ line are distinct, each one of them may correspond to a Dirac point $\mb k$. 
The energy dispersion near the touching point is obtained by expanding to first order in small momenta $\mb q=q(\cos\theta,\sin\theta)$
\begin{equation}
f(\mb k+\mb q)=q(\pd_{k_x}h_1\cos\theta-i\pd_{k_y}h_2\sin\theta)+O(q^2),
\end{equation} 
where on the $T$ line [$\mb k=(k,0)$], we used the property $\pd_{k_y} h_1=\pd_{k_x} h_2=0$. 
Let us define the velocities $c_x=\pd_{k_x} h_1$ and $c_y=\pd_{k_y} h_2$. 
A band touching point $\mb k$ is a Dirac point, 
if both velocities are nonvanishing at $\mb k$, $c_{x}\ne 0 \ne c_{y}$.
More quantitatively $c_x$ and $c_y$ read as
\begin{eqnarray}\label{velocitya}
c_x(x)&=&\mp 2\sqrt{3}\sqrt{1-x^2}[p(x)+(x+\frac{1}{2})p'(x)],\\
c_y(x)&=&-3[p(x)-(x+\frac{1}{2})r(x)],
\end{eqnarray}
where the sign $\mp$ of $c_x$ corresponds to band touchings associated with the $\pm\mb K$ valley, and the polynomial $r(x)$ is given by
\begin{eqnarray}\label{velocityb}
r(x)&=&16t_8 x^3+4(t_4-t_8)x^2+4(t_4-4t_8)x\notag\\
&&+1+t_3-6t_4+2t_7+5t_8.
\end{eqnarray}
To simplify the above equations, it is opportune to study separately the velocities for band touchings 
at the $\pm\mb K$ points ($x=-1/2$), and eventual solutions away from the $\pm\mb K$ points 
on the $T$ line for $x\ne -1/2$ and $p(x)=0$. Let us take the band touchings only at the $\mb K$ valley, 
knowing that the $c_x$ changes sign at the opposite valley. The corresponding velocities read
\begin{eqnarray}\label{velocityc}
x&=&-\frac{1}{2}:\quad
c_x=c_y=-\frac{3}{2}(1-2t_3-t_4+5t_7+4t_8),\notag\\ \\
x&\ne& -\frac{1}{2}:\quad
\begin{cases}
c_x=-2\sqrt{3}\sqrt{1-x^2}(x+\frac{1}{2})p'(x)\notag\\
c_y=3(x+\frac{1}{2})r(x)
\end{cases}.
\end{eqnarray}
The above equations indicate that the two velocities are equal in magnitude and eventual 
Dirac points will have isotropic cones at $\pm\mb K$. Also note that a merging of Dirac points in the $\mb K$ 
valley creates an energy dispersion of higher order in $q$ and this is equivalent to vanishing of the velocities to $c_{x,y}|_{\mb K}=0$.
At band touchings different from $\pm\mb K$ one can use the condition $p(x)=0$ to simplify the expression of the $r(x)$ polynomial: 
\begin{equation}\label{rofx}
r(x)=2(1-x)[4(t_7-t_8)x^2+4(t_7-t_8)x+t_3-2t_4+t_8].
\end{equation}
At time-reversal points $\Gamma(x=1)$ and $M(x=-1)$ the velocity $c_x$ is always zero. The $\Gamma$ point (center of the BZ) 
is the band bottom and presents an isotropic energy dispersion; therefore $c_y$ vanishes together with $c_x$ 
[as seen from Eq.~(\ref{rofx})]. In contrast, at the $M$ point, $c_y$ is not necessarily zero. 
For example, in N3 graphene this allows for $M$ band touchings with linear dispersion in $k_x$ and quadratic in $k_y$. 
These semi-Dirac points correspond to a merging of two Dirac points with opposite chirality.

If {\em all} the band touchings are Dirac points, then their chirality~(\ref{chirality}) follows from Eqs.~(\ref{velocityc}): 
\begin{eqnarray}\label{chiralityb}
x&=&-\frac{1}{2}:\quad\chi(\pm\mb K)=\pm 1,\notag\\
x&\ne& -\frac{1}{2}:\quad\chi(\pm\mb k_i)=\mp\sgn\big[ p'(x)r(x)\big],
\end{eqnarray}
where $\pm\mb k_i$ denote the position of the additional Dirac points associated with the $\pm\mb K$ valley.
The next sections exemplify the above theory to the concrete cases of N3 and N4 graphene.

\subsection{Dirac points and merging for N3 graphene}
\begin{figure}[t]
\centering
\includegraphics[width=0.9\columnwidth]{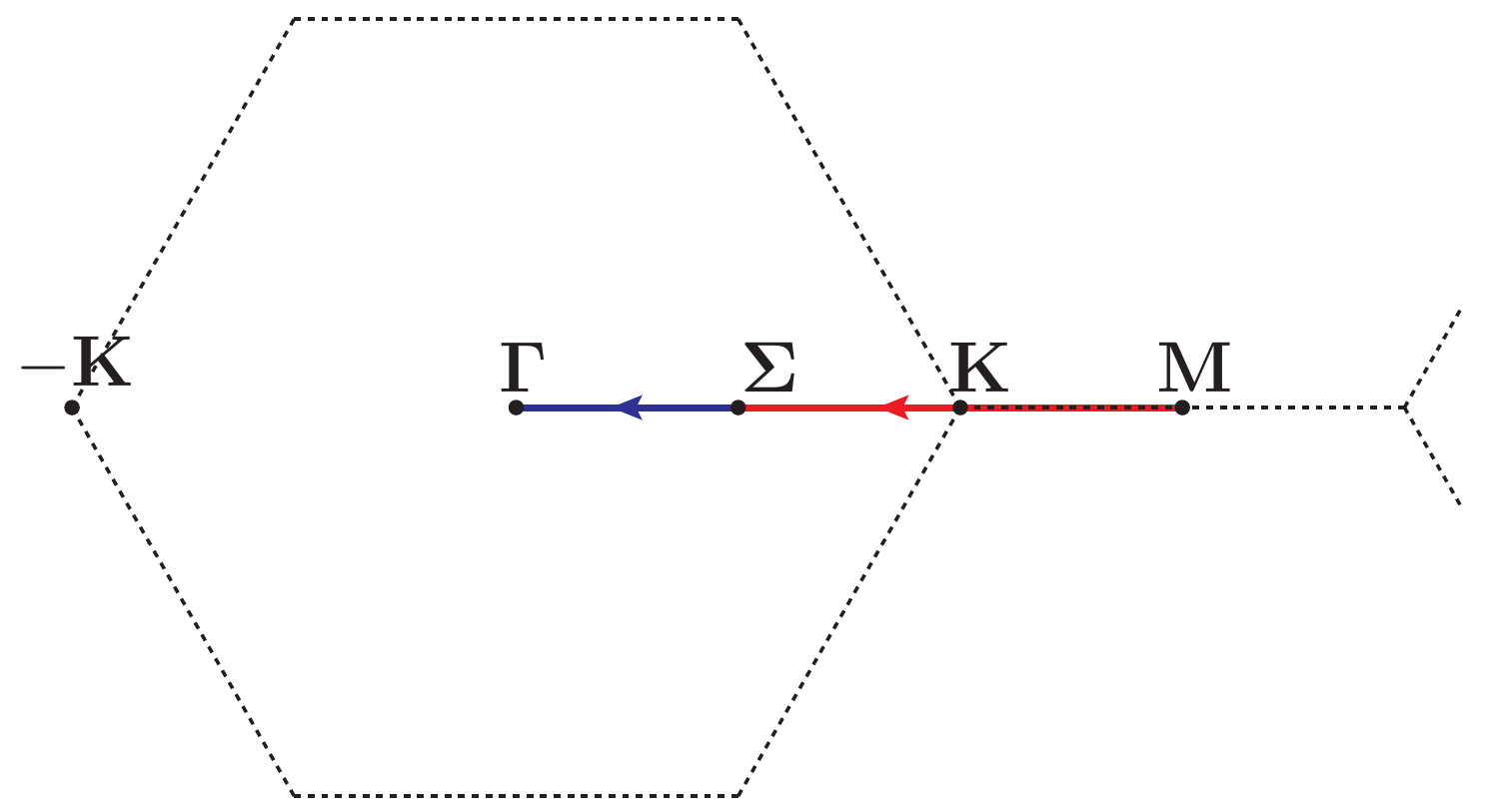}
\caption{(Color online) Evolution in BZ of a satellite Dirac point in N3 graphene on the high-symmetry $T$ line: 
$\Gamma\text{-}\mb K\text{-}M$. The evolution of the satellite point is represented in blue when $t_3$ varies from $-\infty$ to $-1$ 
and in red when $t_3$ varies from $1/3$ to $\infty$.}
\label{fig:move}
\end{figure}

The isotropic N3 graphene was already investigated in Ref.~\onlinecite{Bena}.
Here the presence of a sufficiently strong $t_3$ 
hopping integral was shown to produce three more \textit{satellite} band-touching points orbiting around each regular Dirac point ($\pm\mb K$). 
Indeed solving Eq.~(\ref{points}) (with $t_4=t_7=t_8=0$), it follows that in addition to solution $x=-\frac{1}{2}$ (at $\pm\mb K$), 
there is a solution $x=\frac{t_3-1}{2t_3}$ which may 
give rise to up to six touching points at 
\begin{align}
\pm\mb k_1&=\pm k (1,0),&\pm\mb k_2&=\pm k\bigg(-\frac{1}{2},\frac{\sqrt{3}}{2}\bigg),\\
\pm\mb k_3&=\mp k\bigg(\frac{1}{2},\frac{\sqrt{3}}{2}\bigg),&k&=\frac{2}{\sqrt{3}}\arccos\bigg(\frac{t_3-1}{2t_3}\bigg),\notag
\end{align}
where $\pm\mb k_i$ points are associated with the $\pm\mb K$ valley. 
A physically meaningful solution corresponds to $|x|\le 1$ and hence has an existence domain given by
\begin{equation}
t_3 \in (-\infty,-1)\cup (1/3,\infty).
\end{equation} 
For $t_3 \in (-1,-\infty)$ satellite touching points appear at $\Gamma$ ($t_3=-1$) and move along the $T$ line and reach the $\Sigma$ point ($x=1/2$, $t_3=-\infty$),
midway between $\mb K$ and $\Gamma$.
For $t_3 \in (1/3,\infty)$  satellite touching points appear at $M$ ($t_3=1/3$) and move along the $T$ 
line and reach again $\Sigma$ ($x=1/2$, $t_3=\infty$) (see Fig.~\ref{fig:move}). 
For $t_3\ne 1/2$, the satellites are Dirac points away from the regular Dirac points $\pm\mb K$, $x=\frac{t_3-1}{2t_3} \ne -1/2$.
The chirality associated with the three satellite Dirac point $\mb k_{1,2,3}$ in valley $\mb K$ reads
\begin{equation}\label{chiralityN3}
\chi\bigg(x=\frac{t_3-1}{2t_3}\bigg)=-\sgn[t_3(1+t_3)].
\end{equation}
The chirality $\chi$ is always opposite to points associated with the $-\mb K$ valley.

As already emphasized,\cite{Bena} there is a particular value, $t_3=1/2$, that corresponds to a merging of 
three satellite Dirac points with a central regular Dirac point. 
This case is realized when $x=-1/2$ is a double root of the polynomial $h_1(x)$. Here $p(x)=0$ and
 therefore the velocities $c_{x,y}$ vanish simultaneously, indicating the formation of a band touching with a
 higher than linear dispersion. Note, however, that at the merging point, $\chi(x=-1/2)=\sgn(c_xc_y)$ is not well defined. 
Nevertheless, from Eq.~(\ref{chiralityN3}) it is apparent that the satellite points close to merging at $\pm\mb K$ 
have an opposite chirality from the central Dirac point $\chi(\pm\mb K)=\pm1$. Then the sum rule dictates that the 
chirality at the merging point is the sum of chiralities over the colliding Dirac points. 
At $\pm\mb K$ merging this yields $\chi(t_3=1/2)=\pm(1-3)=\mp 2$.\cite{Volovik,Volovik1}

The chirality of the merging point can be equally determined by expanding the energy dispersion at $\pm\mb K$. 
It suffices to find it at $\mb K$, knowing that time-reversal symmetry demands opposite chirality at $-\mb K$. 
Expanding at $t_3=1/2$ in small momenta $\mb q=q(\cos\theta,\sin\theta)$ it follows that
\begin{equation}
f(\pm\mb K+\mb q)=\frac{9}{8} q^2 e^{\pm 2i\theta}+O(q^3). 
\end{equation}
This indicates that the band touching at the merging of all the Dirac points in a valley has a quadratic dispersion and a topological charge of $\mp 2$ in valley $\pm\mb K$.


\subsection{Dirac points for N4 graphene}
\label{subsec:N4graphene}
\begin{figure}[t]
\begin{overpic}[width=\columnwidth]{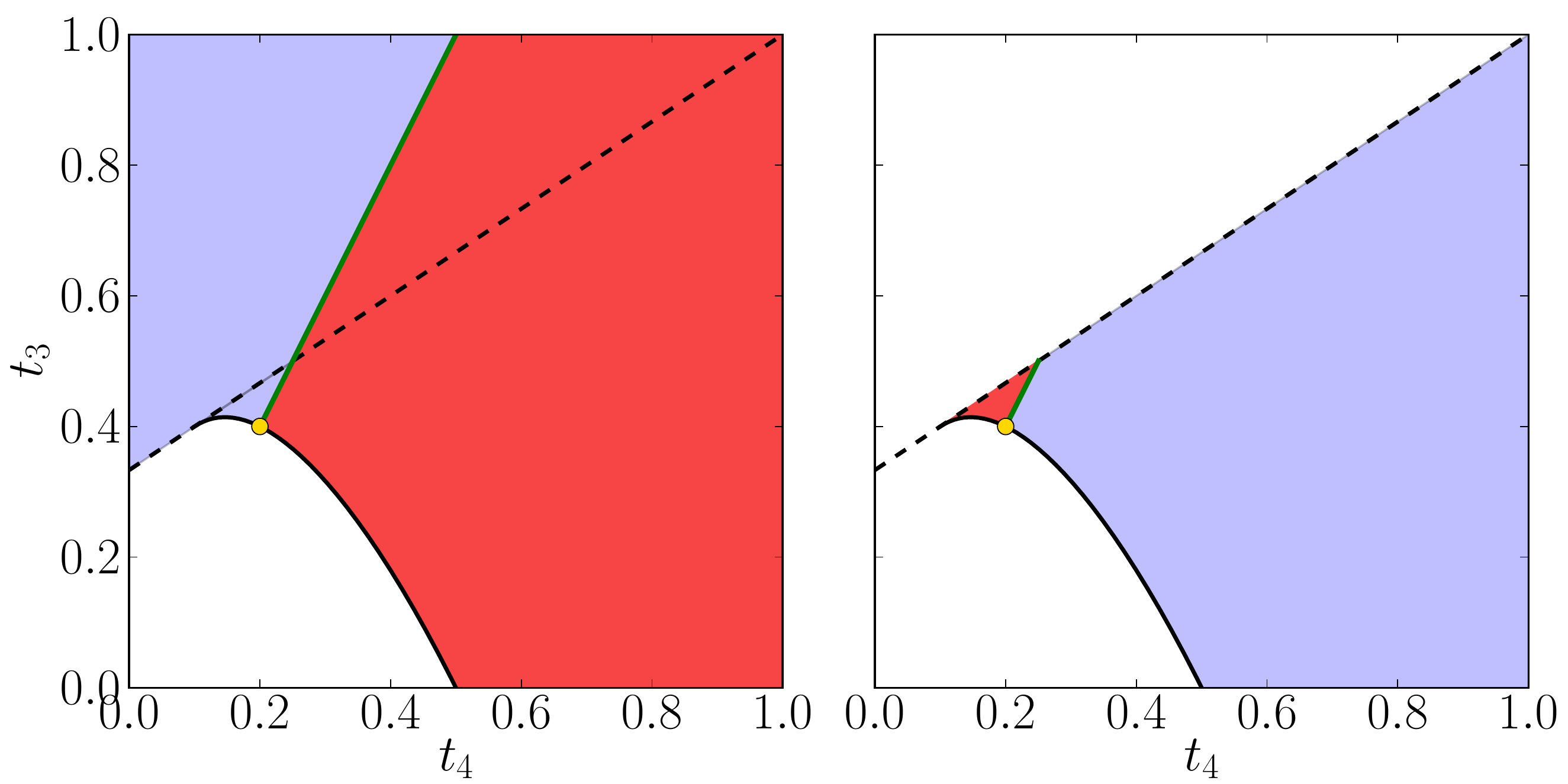}
\put(8.5,44){(a)}
\put(56,44){(b)}
\end{overpic}
\caption{(Color online) (a) $|x_-| \le 1$. (b) $|x_+| \le 1$. 
Existence domains and corresponding chirality of solutions $|x_{\pm}| \le 1$ in $(t_4,t_3)$ parameter space and in the $\mb K$ valley. 
The region with positive (negative) chirality is represented in red (blue). The green line $t_3=2t_4$ 
where chirality changes is associated with the existence of Fermi lines instead of Dirac points. 
The supermerging point $t_3=\frac{2}{5}$ and $t_4=\frac{1}{5}$ at the intersection of the $t_3=2t_4$ line with the 
domain border curve $t_3=2(\sqrt{t_4-t_4^2}-t_4)$ is indicated in yellow.
}
\label{fig:existenceN4}
\end{figure}

\begin{figure}[t]
\centering
\includegraphics[width=\columnwidth]{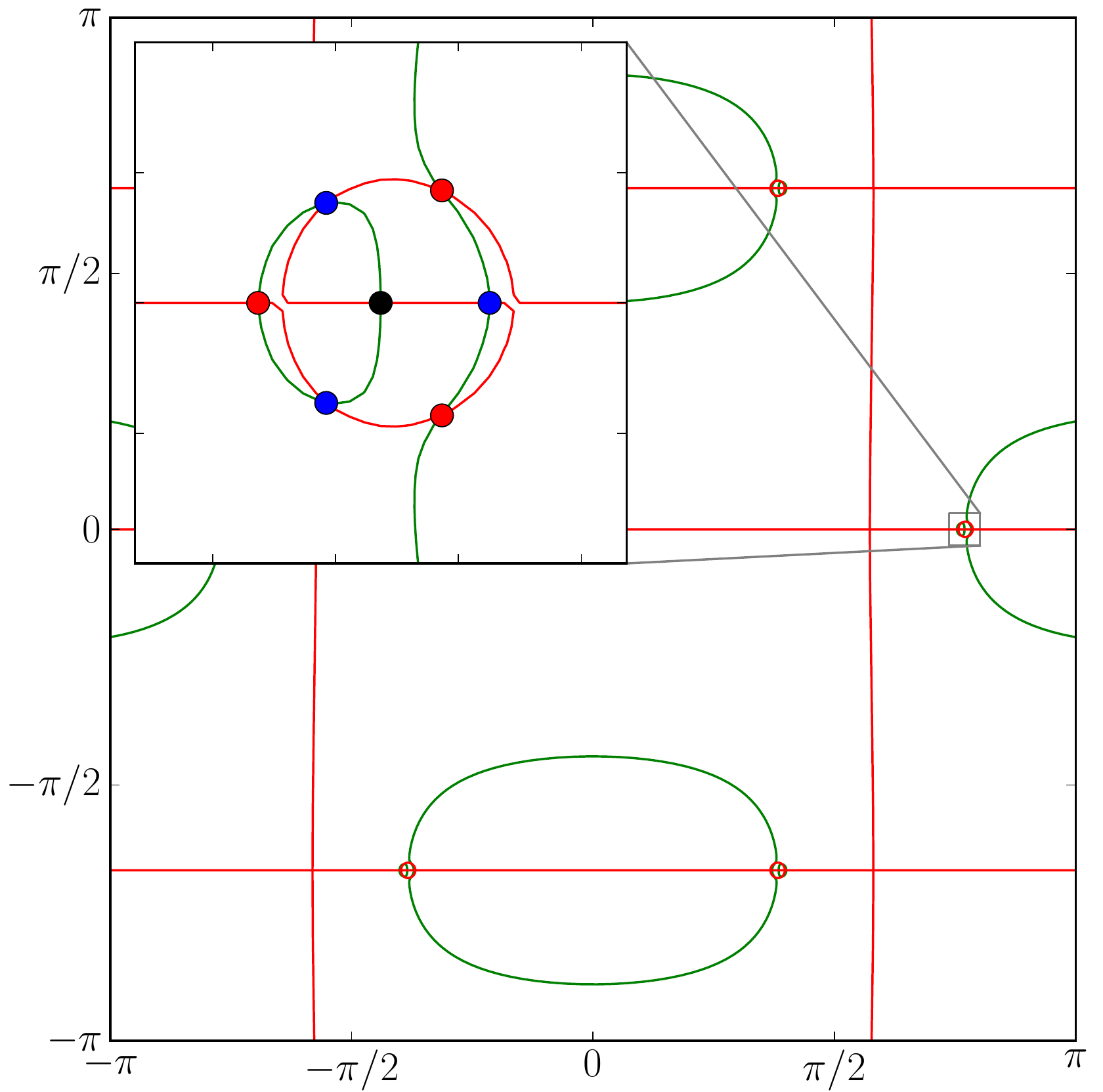}
\caption{(Color online) The zero lines of $h_1(\mb k)=0$ (in green) and $h_2(\mb k)=0$ (in red) for N4 graphene.
A small perturbation $(+0.001)$ of $t_4$ at the merging point $t_3=2/5$, $t_4=1/5$ 
creates six Dirac points around the stable Dirac point $\mb K$. In the inset there is a zoom around $\mb K$. 
The Dirac points are represented by full circles, $\bullet$; there is a central $\mb K$ Dirac point in black, 
and two sets of satellite Dirac points, in blue and red.
}
\label{fig:7dirac}
\end{figure}

For N4 graphene, solving Eq.~(\ref{points}) yields, besides the solution $x=-\frac{1}{2}$ (at $\pm\mb K$), 
two additional solutions $x_{\pm}=-\frac{1}{4t_4}[t_3\pm(t_3^2+8t_4^2+4t_4t_3-4t_4)^{1/2}]$ ($x_+\le x_-$) such that 
there are up to seven band-touching points per valley.
More quantitatively, for $0\le t_3,t_4 \le 1$, one obtains 
the existence domains for additional solutions when $|x_\pm|<1$.
Explicitly, $|x_+|<1$ for
\begin{equation}
t_4 \ge \frac{1}{10}\quad \text{and}\quad 2(\sqrt{t_4-t_4^2}-t_4) \le t_3 \le \frac{1+2t_4}{3}.
\end{equation}
Similarly, $|x_-| \le 1$ holds for
\begin{eqnarray}
\bigg(t_4 &\le& \frac{1}{10}\text{ and }t_3 \ge \frac{1+2t_4}{3}\bigg)
\text{ or } \notag\\
\bigg(t_4 &\ge& \frac{1}{10}\text{ and }t_3 \ge 2(\sqrt{t_4-t_4^2}-t_4)
\bigg).
\end{eqnarray}
The existence domains are represented graphically in Fig.~\ref{fig:existenceN4}.
Note that the two solutions coexist when $t_3\le \frac{1+2t_4}{3}$. In the coexistence region one can generally expect to have seven Dirac points per valley (see Fig.~\ref{fig:7dirac}).

In their existence domain, Eq.~(\ref{chiralityb}) determines the chirality in the $\mb K$ valley
\begin{equation}\label{chiralityN4}
\chi(x_{\pm})=\pm \sgn (t_3-2t_4).
\end{equation}
However, the chirality information is exact when the solutions $x_\pm$ stand for Dirac points. 
The model presents a rich phenomenology and the investigation of the solutions indicates that for
 particular parameters there are also band touchings different from the simple Dirac point case.

Remember that each solution $x_\pm$ stands for a triplet of solutions at each valley. 
Then there are different scenarios for the behavior of Dirac points. There are cases similar 
to the N3 graphene where there is a single triplet of solutions merging to the central Dirac point 
to yield a point with high-energy dispersion. There are cases where the two triplets merge with one 
another to yield a new triplet of band touchings with quadratic dispersion in one direction and linear in the other.
 There is also a unique supermerging point where all Dirac points in a valley merge. A completely new feature to 
the phenomenology of band touchings in N4 graphene is the formation of Fermi lines for specific values of parameters.

The first case is that of a line in parameter space where only one triplet given by the $|x_\pm|$-solutions 
merges with the central Dirac point. These are obtained under the condition that either $x_+=-1/2$ or $x_-=-1/2$,
\begin{equation}
x_{\pm}=-\frac{1}{2}\iff t_3=\frac{1-t_4}{2},\quad t_3\lessgtr 2t_4.
\end{equation}

\begin{figure}[t]
\includegraphics[width=\columnwidth]{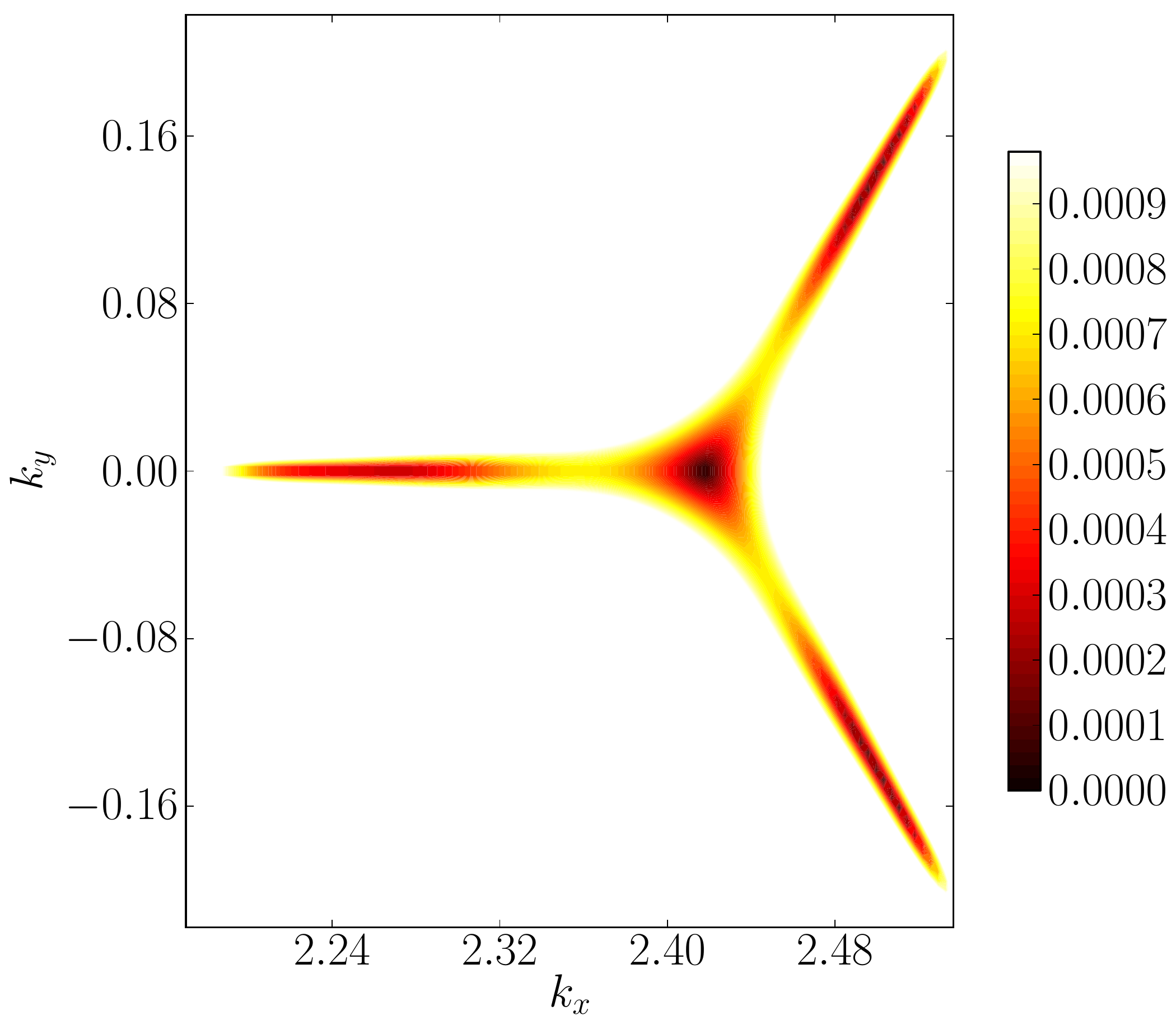}
\caption{Cross section slice through the energy dispersion of the conduction band near zero energy. 
A triplet of semi-Dirac points is formed around the central Dirac point at $\mb K$. On the parameter
 line $t_3=2(\sqrt{t_4-t_4^2}-t_4)$ (here with $t_4=0.25$) the triplets of satellite points merge 
on the high-symmetry lines to form the semi-Dirac points.}
\label{fig:triptrip}
\end{figure}

A different case is that of the triplet satellite Dirac points merging two by two to form semi-Dirac points, 
i.e.\ band touchings with quadratic dispersion in the direction of merging and linear in the direction perpendicular 
to it.\cite{Hasegawa2006, Dietl,Wunsch2008,Prado2009, Banerjee2009,Tarruell2012} 
They correspond to a scenario where two Dirac points 
with opposite chirality collide. From the condition $x_+=x_-$, 
they are determined on the line $t_3=2(\sqrt{t_4-t_4^2}-t_4)$. 
This case is represented in Fig.~\ref{fig:triptrip}

At the intersection of line $t_3=2t_4$ with the domain border curve $t_3=2[\sqrt{t_4(1-t_4)}-t_4]$ ($t_3=\frac{2}{5}$ and $t_4=\frac{1}{5}$),
 there is a supermerging point where there is a unique band touching per valley that can be understood as a collision of all additional Dirac points 
into the central ($\pm\mb K$) one. This point in parameter space has a topological charge given by the sum of all Dirac point chiralities. 
Because of the cancellation of the triplet charges, the final point will have a charge $\pm 1$ in the valley $\pm\mb K$.
Expanding in small momentum $\mb q=q(\cos\theta,\sin\theta)$ around the supermerging point at $\pm\mb K$ yields an effective $f$ function in the $\mb K$ valley,
\begin{equation}
f(\pm\mb K+q)=-\frac{27}{40}q^3 e^{\mp i\theta}+O(q^4).
\end{equation}
This result reinforces the sum rule calculation by showing a band touching with cubic dispersion, but with a low topological charge $\pm 1$ at $\pm\mb K$. 

\begin{figure}[t]
\includegraphics[width=\columnwidth]{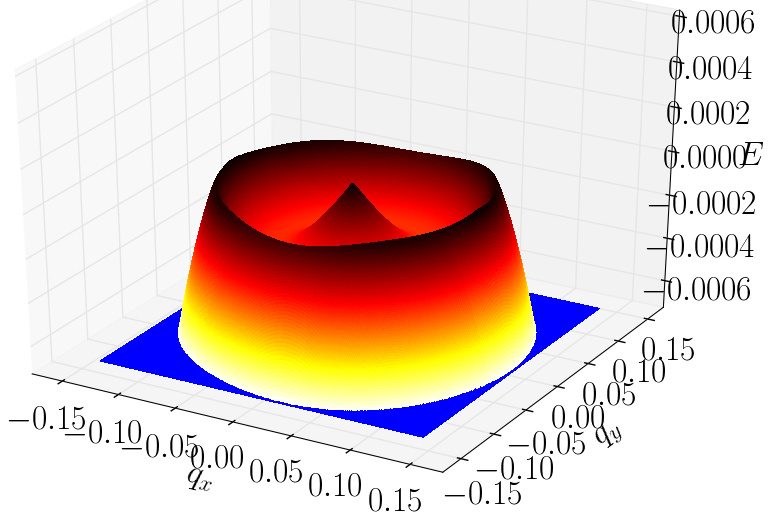}
\caption{
An expansion in small momentum $\mb q$ around the $\mb K$ point of N4 graphene illustrates 
the formation of Fermi lines (lines of zeroes for the energy dispersion) around the regular Dirac point $\mb K$ in graphene, on the parameter line $t_3=2t_4$. 
The hopping parameters are chosen near the supermerging at $t_3=2/5+2\delta$ and $t_4=1/5+\delta$ with $\delta=0.001$.
}
\label{fig:fermiline}
\end{figure}

Finally, there is a phenomenologically new situation that is absent in the previously studied N3 graphene. 
Note that on line $t_3=2t_4$ the chirality~(\ref{chiralityN4}) is zero even though there are nondegenerate solutions $x_-\ne x_+$, away from the supermerging. 
This case corresponds to the existence of closed Fermi lines in the Brillouin zone that link the two solutions. 
Hence, in contrast with the cases studied until now,
 here the energy dispersion exhibits a line of zeros outside the $T$ line. 
One of the cases is represented in Fig.~\ref{fig:fermiline}, where the $x_\pm$ solutions are connected 
 by a Fermi line. However, even for a vanishing $x_+$ solution, Fermi lines subsist and link band touchings associated only to $x_-$ solution. 
Aside from these numerical observations of the Fermi line at $t_3=2t_4$, it remains a daunting task to analytically solve for general 
solutions away from the high-symmetry lines. 
However, one can investigate analytically the peculiarity of this case by considering the behavior 
 of the energy dispersion near the $T$ lines. The absolute value of the energy for $t_3=2t_4$ is 
\begin{equation}
E=|4t_4x^2+4t_4xy-4t_4+1|\sqrt{4x^2+4xy+1},
\end{equation}
where $x=\cos(\sqrt{3}k_x/2)$ and $y=\cos({3k_y/2})$.
It is immediate to verify that the derivatives in the $k_y$ direction for a zero-energy solutions $x_\pm$ on the $T$ line $k(1,0)$ 
vanish at all orders. This indicates that the solution $x_\pm$ are not longer pointlike band touchings, but extend as Fermi lines in the $k_y$ direction.

\subsection{Supermerging at \texorpdfstring{$\pm \mb K$ in N$n$}{+/-K in Nn} graphene}

In the two preceding sections it was shown that for N3 and N4 graphene it is possible to adjust the parameters $t_3,t_4$
so that for each valley all the additional touching points merge with the usual Dirac points at $\pm\mb K$ (a supermerging point).
This means that $x=-1/2$ is a double (respectively, triple) zero of $h_1$ for N3 graphene, $t_3=1/2$ (N4 graphene, $t_3=2/5,t_4=1/5$).
The possibility of finding a set of parameters $t_n$ for which all the additional touching points merge with the usual Dirac points at $\mb K$ 
appears to be valid for all N$n$ graphene and relies essentially on the fact that the polynomial $h_1(x)$ is of a degree
equal to or less than the number $n-1$ of free parameters $t_n$. 
[More precisely, it can be proven that a model with hopping terms at a {\em chemical} distance $m$ will result in polynomial $h_1(x)$ of maximum order $m$.] 
Note that because the number of free parameters grows faster than the degree of the polynomial
 there are no longer {\em unique} supermerging points for graphene N$m$, with $m>7$.
At this supermerging the components $c_{x,y}$ vanish and therefore one needs to go beyond a linear expansion to characterize the neighborhood of $\mb K$.
As an example, it was shown in Ref.~\onlinecite{Bena} that for N3 graphene at the supermerging 
one obtains $f(\pm \mb K + \mb q)\simeq q^2 e^{\pm 2i\theta}$ which now identifies a gapless quadratic dispersion, 
with a phase that is understood as resulting from the sum of the respective chirality of all the merging Dirac points. 
Similarly, for N4 graphene, it followed that $f(\pm\mb K+\mb q)\simeq q^3 e^{\mp i\theta}$. 
The location of the unique supermerging band touching and their associated topological charge are given in Table~\ref{tab:super}. 
Note that the energy dispersion of supermerging band touchings has a higher than linear dispersion. 
However, the topological charge of the converging triplets of satellite points is alternating and hence the resulting topological charge remains low. 

Finally, note that the above scenario of a unique supermerging together with an alternating $\pm1$ and $\mp2$ topological charge at $\pm\mb K$ (see Table~\ref{tab:super}) is not generally valid in N$n$ graphene, and in fact it already breaks down in the N8 model.
For N8 graphene, the supermerging is no longer unique, 
but becomes a line in $(t_3,t_4,t_7,t_8)$ parameter space. Nevertheless, Eq.~(\ref{points}) implies 
that in N7 and N8 graphene there is the same number of satellite Dirac points per valley, because $p(x)$ 
remains a third order polynomial. An expansion near the supermerging point for N7 graphene (see Table~\ref{tab:super}) yields in $\mb K$ valley
\begin{equation}
f(\mb K+\mb q)=\frac{27}{64}\pi^*\big[\pi^3-12t_8(\pi^3-{\pi^*}^3)\big],
\end{equation}
with $\mb\pi=q_x+iq_y$. For vanishing $t_8$, one recovers a topological charge $-2$ 
for the band touching at $\mb K$, at the supermerging in N7 graphene. However, when $t_8$ reaches the critical value $1/12$ the band touching clearly exhibits the topological charge $4$. 
This scenario can be explained by a change in chirality for a triplet of Dirac points in the vicinity of the supermerging line $(1-3+3-3)\to(1-3+3+3)$. 
However the behavior of solutions on the supermerging line in N8 graphene is beyond the scope of the present paper.

The N$n$ graphene model was shown to exhibit more than one touching point in each valley. Now it remains to answer the question whether large Chern number phases become possible when gapping them with a Haldane mass. As long as that the position of the band touchings and their respective chirality is known, determining the topological phase diagram is within analytical grasp. 

\begin{table}
\caption{Supermerging characteristics at $\mb K$. The function $f$ from the effective low-energy Hamiltonian
$H_{\rm eff}=\frac{1}{2}\sigma_+f+{\rm H.c.}$ is written as a function of small momenta $\pi=q_x+iq_y$ and up to a multiplicative constant which is neglected.}
\begin{center}
\begin{tabular}{c c c c c c c}
\hline\hline
\multirow{2}{*}{Graphene}& \multicolumn{4}{c}{Supermerging} & \multirow{2}{*}{$f(\mb K+\mb q)$} & \multirow{2}{*}{Charge}\\
& $t_1$ & $t_3$ & $t_4$ & $t_7$ & & \\
\hline
N1 & 1 & 0 & 0 & 0 & $\pi^*$ & $1$ \\ 
N3 & 1 & $1/2$ & 0 & 0 & ${\pi}^2$ & $-2$ \\
N4 & 1 & $2/5$ & $1/5$ & 0 & ${\pi^*}^2\pi$ & $1$ \\
N7 & 1 & $7/12$ & $1/4$ & $1/12$ & ${\pi^*}\pi^3$ & $-2$ \\
\hline\hline
\end{tabular}
\end{center}
\label{tab:super}
\end{table}

\section{Chern number phase diagram for the long distance hopping Haldane model}\label{sec:three}

The Haldane model is built on the hexagonal lattice for N1 graphene by adding N2 (intrasublattice) hopping $t_2$, 
such that when hopping is performed clockwise in the unit cell an electron gains a phase $\phi$. 
However there is no net magnetic flux in the unit cell.  
The N2 hopping term leads to two contributions of respective form $h_0({\bm k}) \sigma_0$ with  $h_0({\bm k}) =h_0(-{\bm k})$
and  $h_3({\bm k}) \sigma_3$ with  $h_3({\bm k}) =-h_3(-{\bm k})$. 
These two contributions break chiral symmetry, but do not break inversion symmetry. 
The first contribution breaks particle-hole symmetry, while the second breaks  time-reversal symmetry. 
As noted before, the first contribution does not weight on the Chern number calculation and therefore can be discarded, provided the second contribution produces the necessary 
mass term from Eq.~(\ref{chno}) that gaps the Dirac points. 
That is to say, the topological properties of each band are unaffected by smooth deformations that preserve a finite direct gap at all momenta.
As we shall see later, the second contribution $h_3\sigma_3$ allows for a Chern phase diagram with only odd (even) 
Chern number phases when added to the N$n$ graphene model. In order to have a 
Chern phase diagram allowing for transition between even and odd  Chern number phases, it is necessary to add a mass term that breaks inversion symmetry.
The simplest such term is of the form $M\sigma_3$ and corresponds to a different on-site potential energy on each sublattice.

The mass term $h_3\sigma_3$ in the original Haldane model breaks time-reversal and inversion symmetry. It reads
\begin{equation}
h_3=M-2t_2\sin\phi\{\sin(\mb k\cdot\mb a_2)-\sin(\mb k\cdot\mb a_1)+\sin[\mb k\cdot(\mb a_1-\mb a_2)]\}.
\end{equation}
When intrasublattice hopping between distant sites is allowed, the generalized mass term reads
\begin{eqnarray}
h_3&=&M-\sum_n 2t^{(n)}\sin(n\phi)\{\sin(n\mb k\cdot\mb a_2)-\sin(n\mb k\cdot\mb a_1)\notag\\
&&+\sin[n\mb k\cdot(\mb a_1-\mb a_2)]\},
\end{eqnarray}
where $n$ is an integer that indicates that hopping takes place between $AA$ or $BB$ sites situated at a distance of $n\sqrt{3}a$. Here will be considered only the first two terms in this expression, corresponding to a hopping across two unit cells (see Fig.~\ref{fig:hexhop} and Table~\ref{tab:haldmass}). The term containing the hopping integral $t_5$ just multiplies the identity Pauli matrix and is neglected. Interesting for the topology of the problems are hoppings along the links where the electrons gain the phase $\phi$. Here only the first two terms in the mass term are considered: $t_2$ and $t_6$.

\begin{table}[t]
\caption{The first hopping integrals $t_n$ contributing to the Haldane mass. The hopping distances are expressed in units of lattice constant.}
\centering
\begin{tabular}{ccc}
\hline\hline
Hopping & Physical distance & Chemical distance \\
\hline
$t_2$ & $\sqrt{3}$ & 2 \\
$t_5$ & 3 & 4 \\
$t_6$ & $2\sqrt{3}$& 4 \\
\hline\hline
\end{tabular}
\label{tab:haldmass}
\end{table}

The goal of this part is to illustrate how gapping the graphene system with $2n$ Dirac points can yield $\mbb Z$ topological phases characterized by a large Chern number (up to $\mc C=\pm n$). 
The following sections investigate cases where different mass term gaps the previously obtained N$n$ graphene. The strategy will be to illustrate the possibility of large Chern phases by considering first the action of $t_2$ Haldane mass on different models of N$n$ in Sec.~\ref{sec:t2}. In Sec.~\ref{sec:t6} it is shown that the addition of $t_6$ terms allows one to further increase the absolute value of the Chern number.

\subsection{\texorpdfstring{$t_2$}{t2} Haldane model}
\label{sec:t2}
\begin{figure}[t]
\centering
\includegraphics[width=\columnwidth]{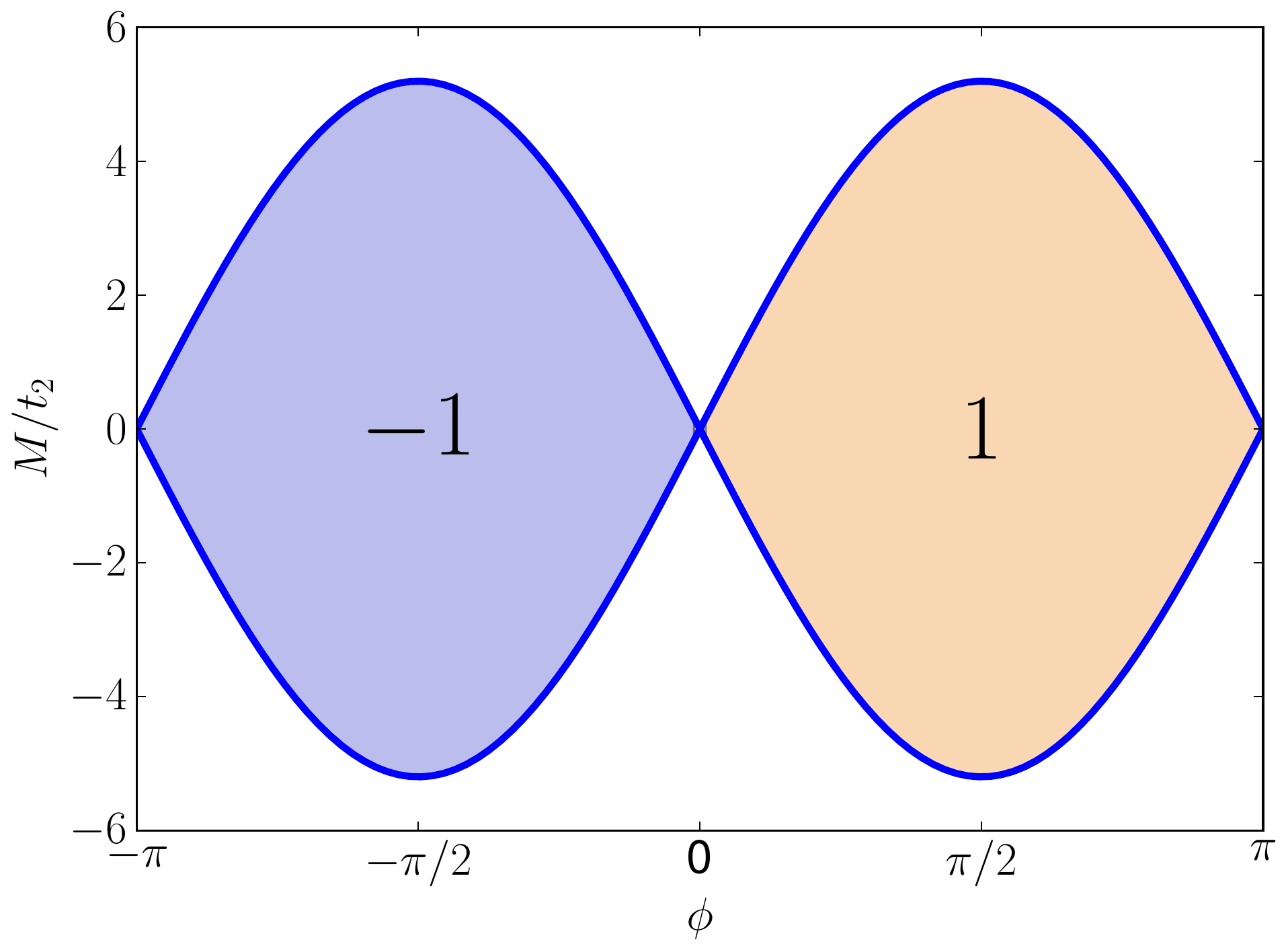}
\caption{(Color online). Chern number phase diagram for the Haldane Hamiltonian as a function of the on-site energy $M$ divided by the hopping integral $t_2$ as a function of the flux $\phi$. The topologically nontrivial insulating phases are color identified and have the topological index denoted inside the respective regions. The topologically insulating regions, $\mc C=0$, are white.}
\label{fig:hald}
\end{figure}

N1 graphene with a hopping $t_2$ constitutes the original Haldane model. 
The phase diagram is obtained by observing that $h_3$ changes sign between the Dirac points ($\mp\mb K$) of graphene. Therefore the Hamiltonian exhibits three topological phases: a trivial insulating phase and two $\mc C=\pm 1$ quantum anomalous Hall (QAH) phases. Equation~(\ref{chno}) yields in this case 
\begin{equation}
\mc C=\frac{1}{2}(\sgn\mathcal M_--\sgn\mathcal M_+),
\end{equation}
where $\mathcal M_{\pm}=M\mp3\sqrt 3 t_2\sin\phi$ is the mass term at $\mp\mb K$. The phase diagram is represented in Fig.~\ref{fig:hald}. The lines $\mc M_\pm=0$ represent topological transition lines where the bulk gap closes at least at one of the $\pm\mb K$ points.

Larger Chern phases become possible when the underlying model is N3 graphene.
Now the mass term takes different values between a regular Dirac point and its satellites. Therefore the topological charges can add up to yield Chern $|\mc C|=2$ phases.

Momentum $\pm\mb k_i$ locates any satellite point of $\pm\mb K$ and, manifestly, the expression for $\chi(\mb k_i)$ holds in the range of existence of separate satellite points.

\begin{figure}[t]
\centering
\includegraphics[width=\columnwidth]{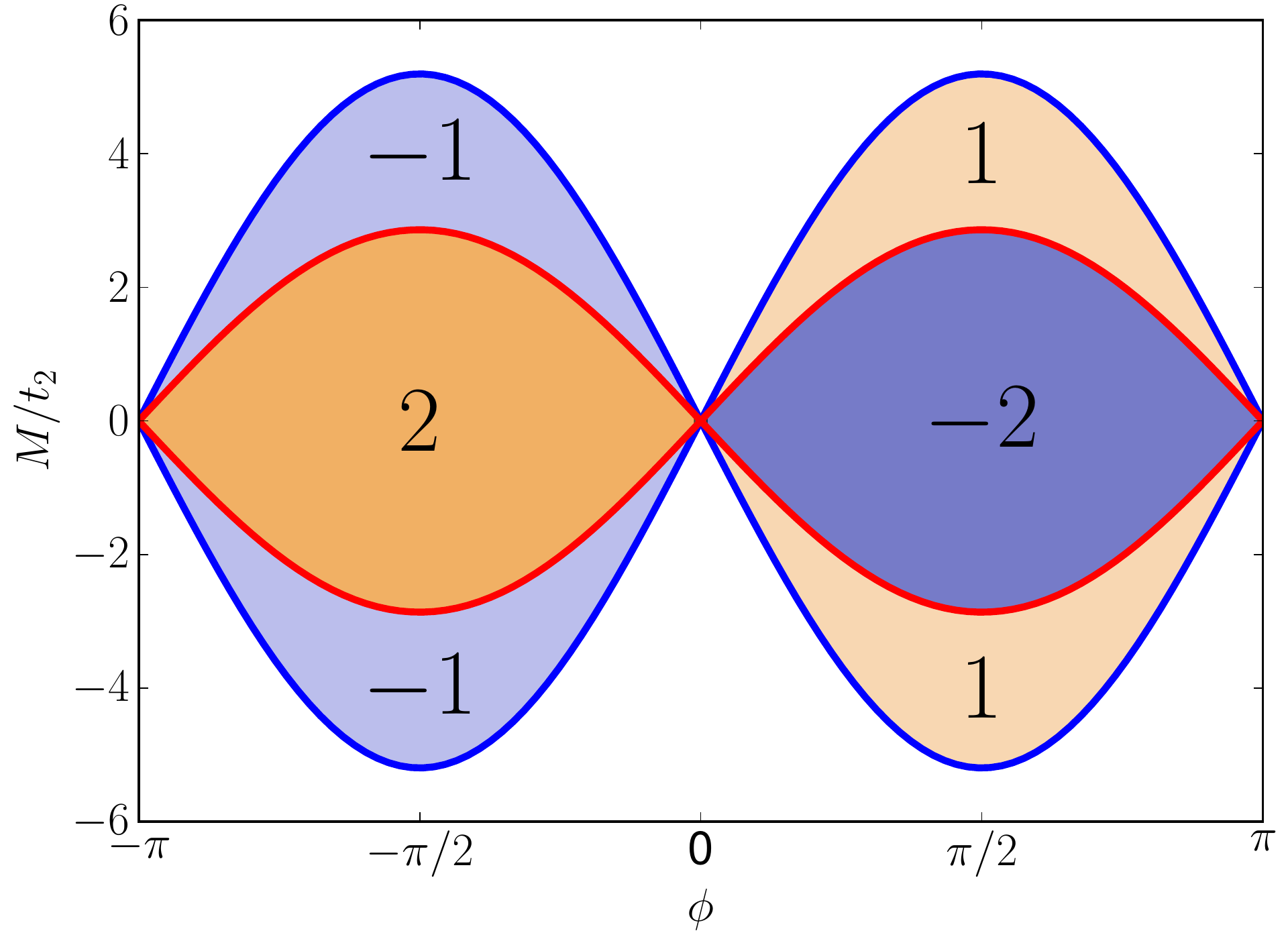}
\caption{(Color online) Chern number phase diagram for the $t_2$ Haldane model on N3 graphene. The hopping parameters are $t_2=1/3$ and $t_3=0.35$ in units of $t_1$.}
\label{fig:modhald}
\end{figure}

Let us define the mass at the regular Dirac points $\mathcal M_{\pm}=h_3\big(\mp(\frac{4\pi}{3\sqrt{3}},0)\big)$. Similarly, the mass at the satellite Dirac points $\mb k\ne\mb K$ is denoted by $m_{\pm}=h_3(\mb k)$ in valley $\mp\mb K$. Then from Eq.~(\ref{chno}) it follows that the Chern number is
\begin{equation}\label{chnofinal}
\mc C=\frac{1}{2}\bigg[(\sgn\mathcal M_- - \sgn\mathcal M_+)-3(\sgn m_- - \sgn m_+)\bigg]
\end{equation}
where the mass of the Dirac points read
\begin{eqnarray}\label{mass}
\mathcal M_{\pm}&=&M\mp 3\sqrt{3}t_2\sin\phi,\notag\\
m_{\pm}&=&M\mp 2\frac{t_2}{t_3}(1+t_3)
\sqrt{1-\bigg(\frac{1-t_3}{2t_3}\bigg)^{\!2}}\sin\phi.
\end{eqnarray}

\begin{figure*}[t]
\centering
\includegraphics[width=\columnwidth]{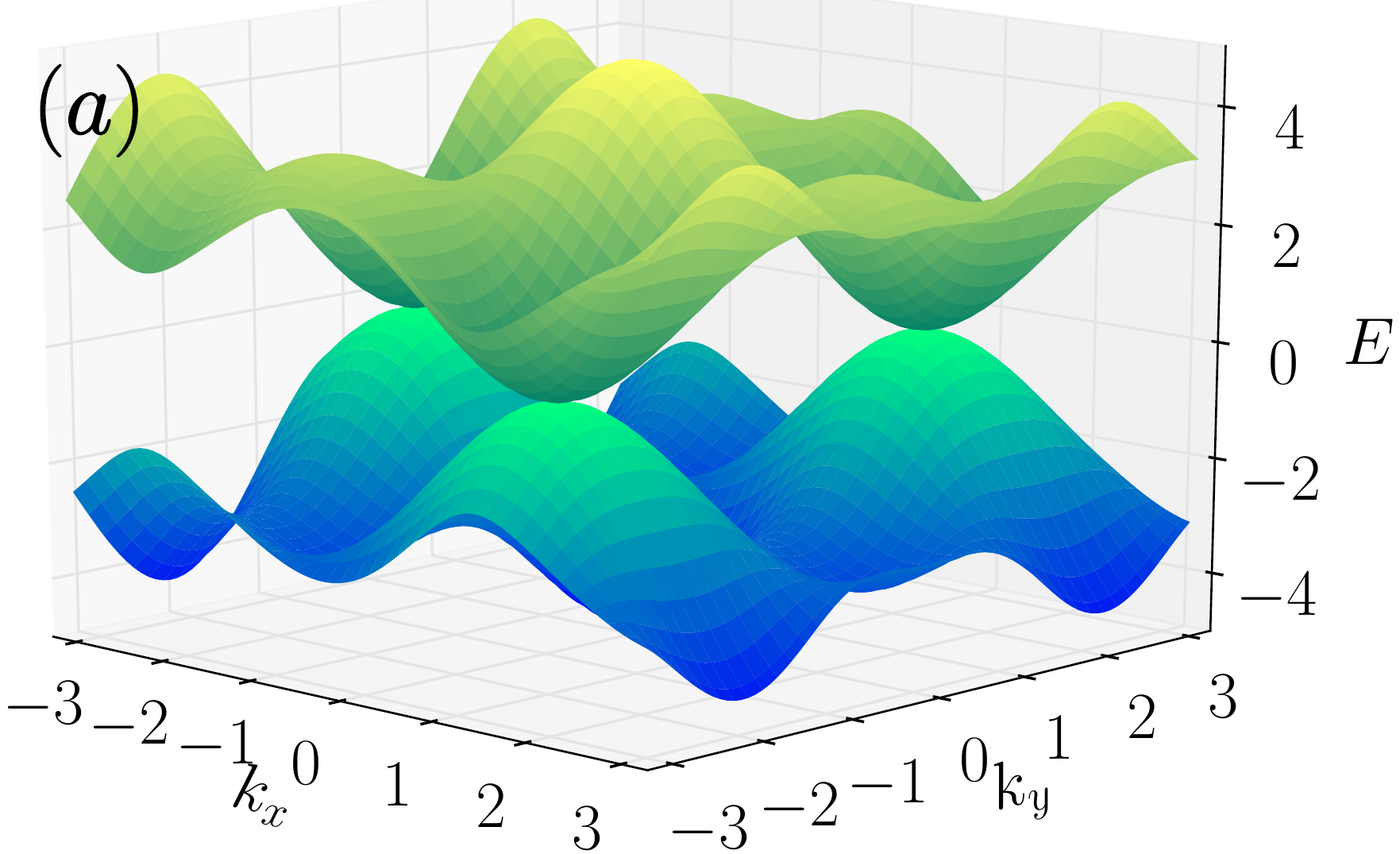}
\includegraphics[width=\columnwidth]{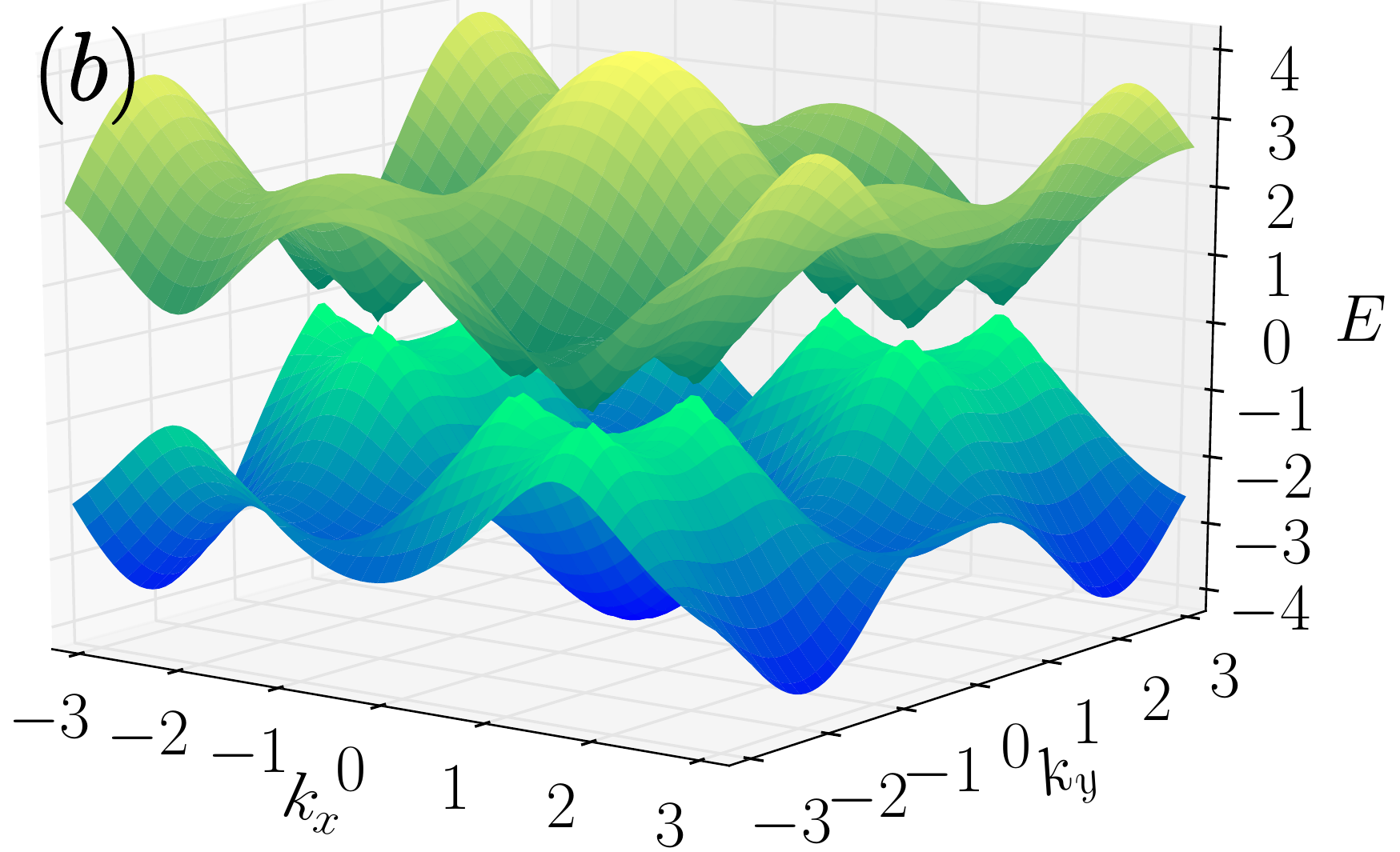}
\caption{(a) Energy dispersion at the topological transition between $\mc C=-2$ and $\mc C=0$ phases at the merging point between the regular $-\mb K$ and its three satellites $-\mb k_i$ in N3 graphene. The energy dispersion in N3 Haldane shows a quadratic band touching at $-\mb K$. The parameters are chosen $\phi=\pi/2$, $M=\sqrt{3}$, $t_2=1/3$ and $t_3=1/2$. (b) Energy dispersion for the N3 Haldane model at the transition between $\mc C=1$ and $\mc C=-2$ phases. The Dirac cones form at the satellite points of $-\mb K$ for $t_2=1/3$, $t_3=0.35$ in units of $t_1$. The change in Chern number by three units is reflected in the presence of three Dirac cones at the topological transition.}
\label{fig:quad}
\end{figure*}

Equation~(\ref{chnofinal}) yields the phase diagram for the system when all eight Dirac points are present. When there are no satellite Dirac points [$t_3\in(-1,1/3)$], the topology of the system is in fact identical to the original system $t_3=0$ and therefore it has the phase diagram in Fig.~\ref{fig:hald}. When $t_3$ is varied to go outside the region $(-1,1/3)$, two phases of higher Chern number develop around the $M=0$ line. For example, from Eqs.~(\ref{mass}), we see that at $M=0$ a regular Dirac point and its satellites will have the same mass. Therefore the Chern number reduces to $\mc C=\sgn\mc M_+-\sgn\mc M_-$. This yields topological phases indexed by $\pm 2$. By increasing $|M|$, one crosses a transition line where the Haldane mass of all satellite points in the system becomes identical, while it remains different for the regular Dirac points. This transition is given by
\begin{equation}
m_{\pm}=0.
\end{equation} 
This region extends up to the the last topological transition line given by $M=\pm 3\sqrt{3}t_2\sin\phi$. In this region the Chern number reduces again to the original case ($t_3=0$) with $\mc C=1/2(\sgn\mathcal{M_-}-\sgn\mathcal{M_+})$. When $M$ is increased even further, all Dirac points are gapped identically and therefore this is the topologically trivial region. In Fig.~\ref{fig:modhald} is represented a typical phase diagram for the case where satellite Dirac points are present. 

Note that at the merging point $t_3=1/2$ the $\mc C=\pm 1$ phases completely vanish, and the phase $\mc C=\pm 2$ would have maximal area delimited by $M=\pm 3\sqrt{3}t_2\sin\phi$. Then at the topological transition from the $|\mc C|=2$ phase to the trivial insulator, there is a quadratic band touching that is represented in Fig.~\ref{fig:quad}(a).

The phase diagram in the N3 Haldane model (Fig.~\ref{fig:modhald}) has the nice feature that it accommodates lines of transition where the Chern number changes by three units. This is realized by the formation of three Dirac points at the topological transition. These band touchings come from the vanishing of the Haldane mass at the three satellite Dirac points previously found in N3 graphene. For example, let us take parameters $t_1=1$, $t_2=1/3$ and $t_3=0.35$ from the phase diagram in Fig.~\ref{fig:modhald}. Then fixing $\phi=\pi/2$, there are two transition points between $\mc C=-2$ and $\mc C=1$ phases near $\pm\mb K$. In particular, near $-\mb K$, the Dirac points form at the satellites where $m_+=0$. The energy dispersion at the topological transition is illustrated in Fig.~\ref{fig:quad}(b).

\begin{figure}[t]
\includegraphics[width=\columnwidth]{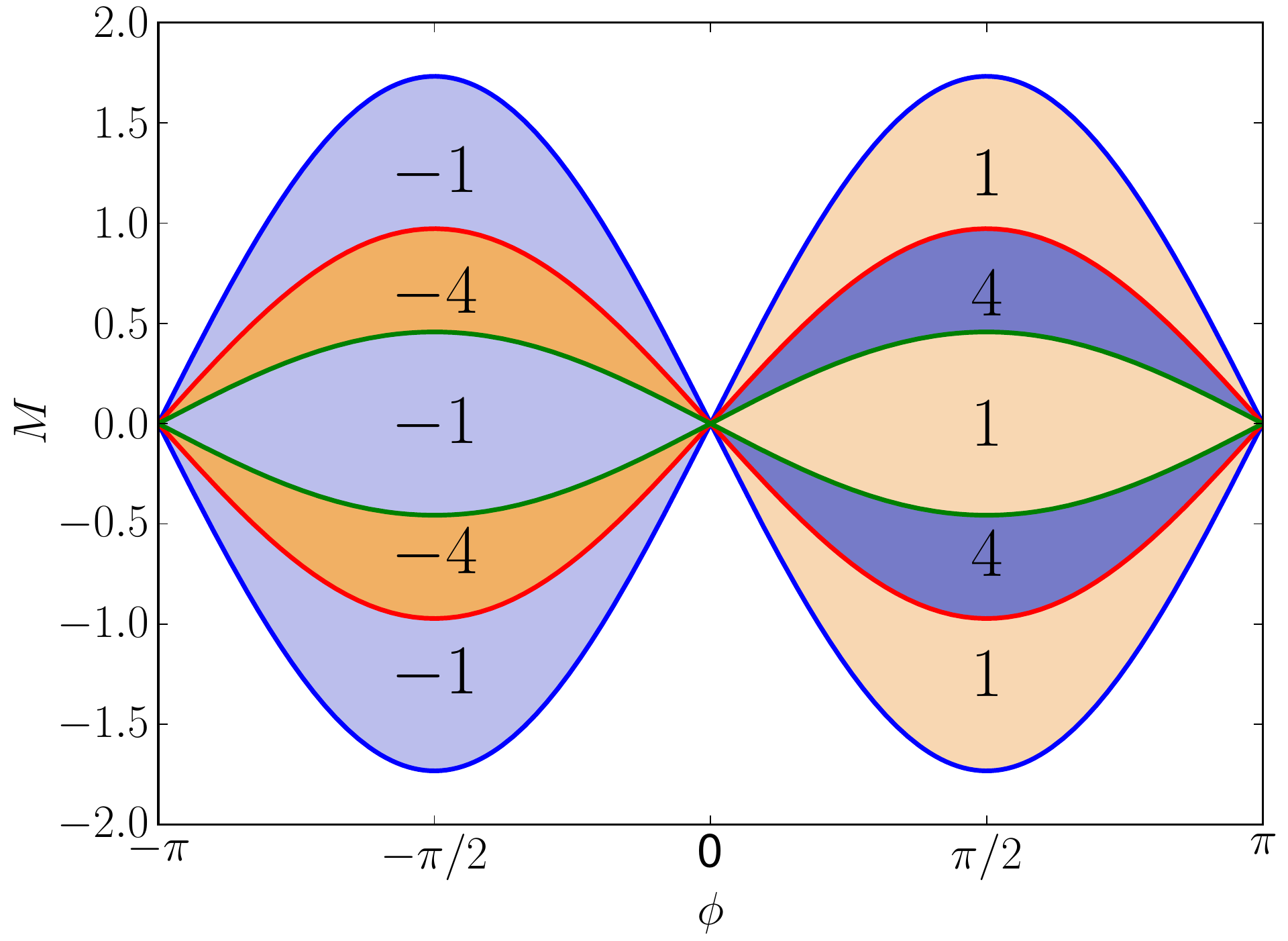}
\caption{Chern number phase diagram showing the existence of 2 sets of satellite Dirac points and large QAH phases in $t2$ Haldane model on N4 graphene. The parameters are $t_1=1$, $t_2=1/3$, $t_3=0.59$ and $t_4=0.4$.}
\label{fig:phdiagt4}
\end{figure}

Similarly one can take as the underlying model the N4 graphene model which contains the $t_4$ hopping. This was shown to produce seven Dirac points per valley. Hence one can expect the presence of larger Chern phases. This is exemplified in Fig.~\ref{fig:phdiagt4}, where a choice of particular parameters yields $|\mc C|=4$ QAH phases. Note the presence of multiple Dirac points is reflected in the phase diagram as a multiplication of transition lines in the $M$ direction for fixed magnetic flux $\phi$ ($\ne 0,\pi$). 

\subsection{\texorpdfstring{$t_6$}{t6} Haldane model}
\label{sec:t6}
The existence of $2n$ Dirac points for a submodel containing only two sigma matrices allows one, in principle, to build topological insulators with Chern phases $\mc C=n$. For the N3 graphene model with eight Dirac points, one can have a large Chern number $\mc C=\pm 4$. To actualize all possible topological phases it is sufficient to add a $t_6$ mass term. It has the effect to produce oscillations in the phase dependent Haldane mass, such that the term changes sign between a regular graphene Dirac point and its satellites in N3 graphene. As expected, all phases are attainable under this modification of the Hamiltonian.

\begin{figure}[t]
\includegraphics[width=\columnwidth]{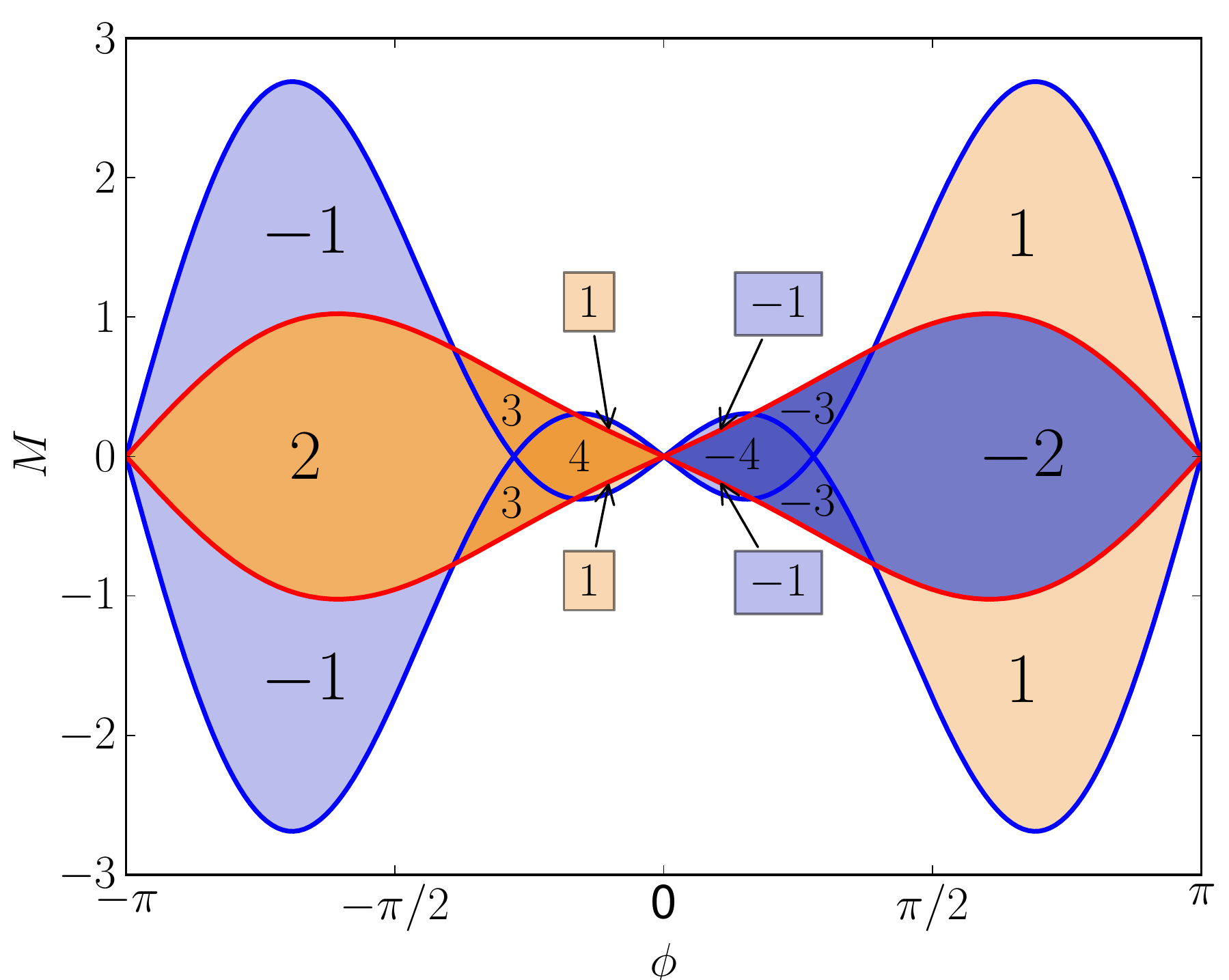}
\caption{All QAH phases possible for N3 graphene with $t_6$ Haldane mass; here the phase diagram for the parameter choice $t_1=1$, $t_2=1/3$, $t_3=0.35$, and $t_6=0.26$ illustrates this point. For $M=0$, the possible Chern phases have only even Chern numbers.}
\label{fig:phdiag4}
\end{figure}

The mass term becomes
\begin{eqnarray}\label{modmass}
h_3&=&M-2t_2\sin\phi\{\sin(\mb k\cdot\mb a_2)-\sin(\mb k\cdot\mb a_1)\notag\\
&&+\sin[\mb k\cdot(\mb a_1-\mb a_2)]\}-2t_6\sin(2\phi)\{\sin(2\mb k\cdot\mb a_2)\notag\\
&&-\sin(2\mb k\cdot\mb a_1)+\sin[2\mb k\cdot(\mb a_1-\mb a_2)]\}.\notag\\
\end{eqnarray}

\begin{figure}[t]
\includegraphics[width=\columnwidth]{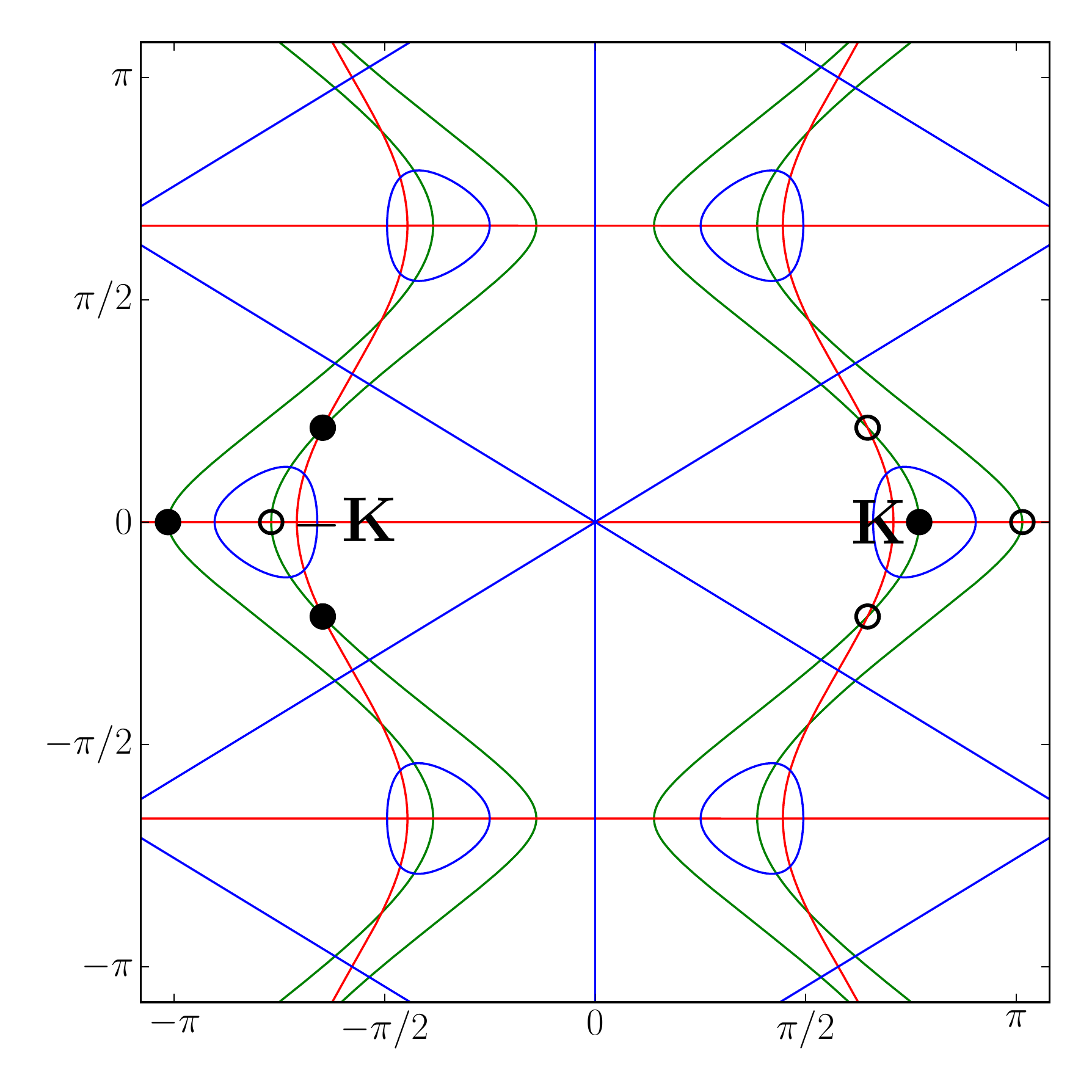}
\caption{(Color online). A Dirac point that is represented by $\bullet$ ($\circ$) has chirality $+$ ($-$). The colored lines represent lines of zeros for $h_1$ (green), $h_2$ (red), and the mass term $h_3$ (blue). The regular Dirac points placed at $\big(\pm\frac{4\pi}{3\sqrt{3}},0\big)$ are gapped by a Haldane mass that has opposite sign. Also the mass term changes sign between the regular Dirac point and its satellites. For parameters $t_1=1$, $t_2=1/3$, $t_3=0.35$, $t_6=0.26$, $M=0$ and $\phi=\pi/8$ the phase is $\mc C=-4$.}
\label{fig:four}
\end{figure}

The new phase diagram is computed by considering the mass term $(\ref{modmass})$ at the eight N3 graphene Dirac points. Then the topological transition lines are given by the zeros of the new mass terms, $\mc M'_{\pm}$ and $m'_{\pm}$ expressed as a function of the previous mass terms from Eq.~(\ref{mass}),
\begin{eqnarray}
\mc M'_{\pm}&=&\mc M_{\pm}\pm 3\sqrt{3} t_6\sin 2\phi\notag\\
m'_{\pm}&=&m_{\pm}\mp 2t_6\sin2\phi(2\sin 2\kappa-\sin 4\kappa),
\end{eqnarray}
where $\kappa=\arccos[(t_3-1)/(2t_3)]$ in the domain of existence of the satellite Dirac points in N3 graphene.

The dependence of the mass term on $\sin 2\phi$ makes possible large Chern number phases $|\mc C|=\pm 4$ by having the mass term changing sign between the regular Dirac cones and its time-reversed one and its own satellites (see Fig.~\ref{fig:four}). When system parameters are varied, the N6 Haldane model can present all Chern phases between $-4$ and $4$. A phase diagram that illustrates this point is represented in Fig.~\ref{fig:four}. The phase diagram was also sampled by numerical integration over the BZ in Eq.~(\ref{intch}) and the results were in agreement.
 
\begin{figure}[t]
\centering
\includegraphics[width=\columnwidth]{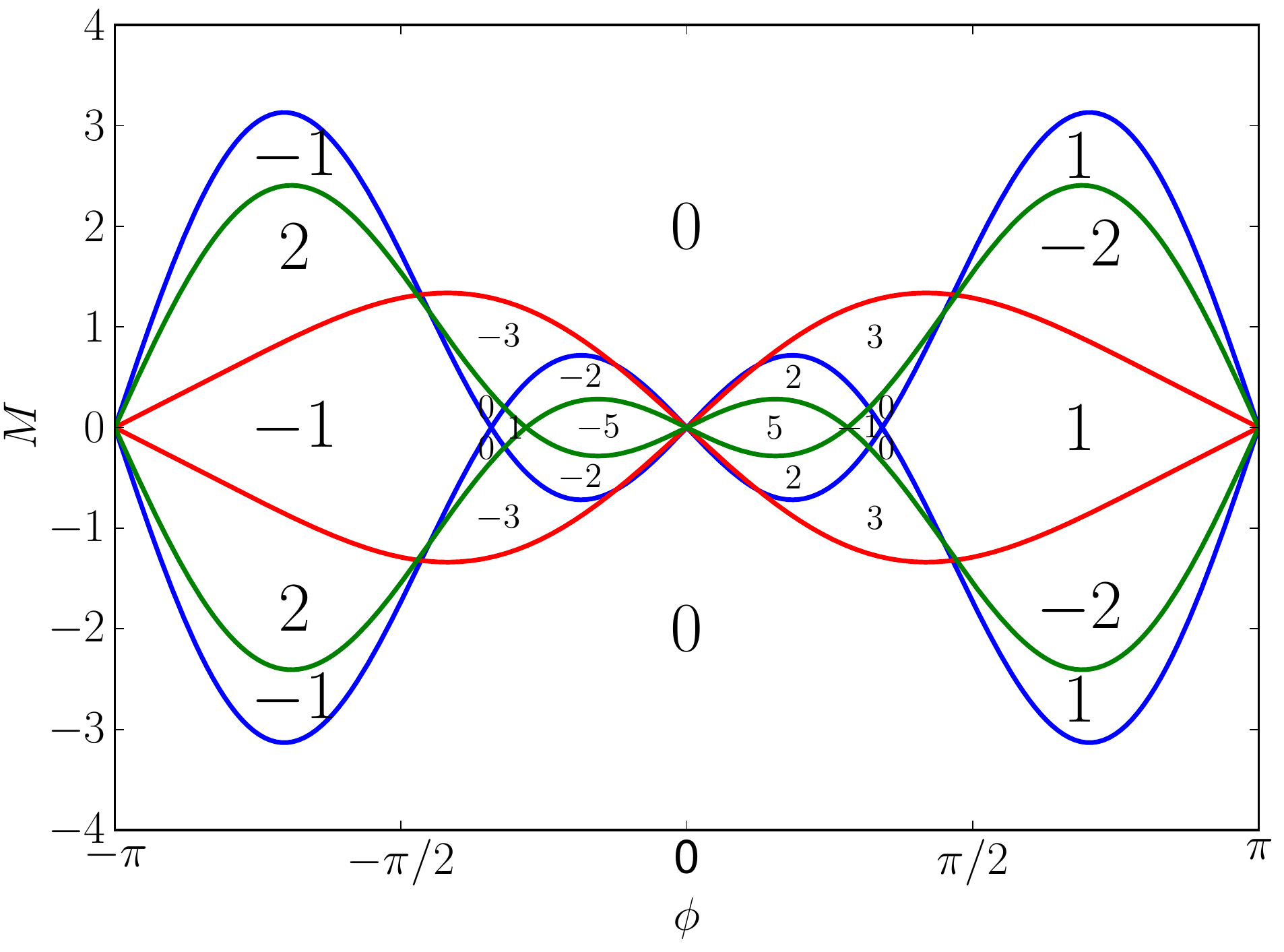}
\caption{Haldane model from N4 graphene with a $t_6$ mass term. Hopping integrals $t_1=1$, $t_2=1/3$, $t_3=0.43$, $t_4=0.3$, $t_6=0.35$. For $M=0$, the possible Chern phases have only odd Chern number.}
\label{fig:phdiagn7}
\end{figure}

Let us consider briefly the case of N4 and N7 graphene by adding, respectively, $t_4$ and $t_7$ hopping terms. With hopping integral $t_1$ fixed as before, there are two free parameters $t_3$ and $t_4$. The parameter space becomes too large to describe analytically the dynamics of the Dirac points and to track at the same time the sign of the mass at the Dirac points. The general thesis,however, remains correct. Larger and larger QAH phases become possible. In the case of N4 graphene there is a maximum of six Dirac points near a $\mb K$ point; for N7 graphene there are nine possible Dirac points per valley. That indicates that with a proper mass term one can have the largest Chern phases $|\mc C|=7$ (in N4 graphene) or $|\mc C|=10$ (in N7 graphene).
In Fig.~\ref{fig:phdiagn7} is represented a Haldane $t_6$ mass on a N4 graphene with QAH phases $|\mc C|\le 5$. It appears that one needs even longer hopping terms in the Haldane mass to realize the largest $|\mc C|=7$ phase.

Note that in all the cases the presence of distant-neighbor hoppings in Haldane mass potentially leads to more bulk gap closings at a given on-site energy $M$ for $\phi$ varying from $-\pi$ to $\pi$. This is reflected in the structure of the phase diagrams, which present oscillations of the topological phase boundaries in the $\phi$ direction. This accounts for the oscillatory nature of the Haldane mass, which can pass more times through zero (as a function of the flux), when it contains strong distant-neighbor hopping terms.

\section{Conclusion}
We have shown that in a graphenelike system adding distant-neighbor hopping integrals leads to the apparition of satellite Dirac points in the spectrum near the regular $\pm\mb K$ points of graphene. The number of additional Dirac points grows as more distant-neighbors are considered. Here, Dirac points up to N7 (next$\times$ 6-nearest-neighbor) graphene model are determined. Each new distant hopping integral between $AB$ sites potentially yields a triplet of Dirac points near $\mb K$ (and because TRI, a triplet at $-\mb K$). For N7 graphene there is a maximum of three triplets of satellites created.

The position of the nodes in the dispersion requires solving a polynomial whose degree grows as more distant neighbors are considered. Analytically, one can hope to determine their position only for a limited number of added neighbors (here N4 graphene). Already, for N4 graphene, the investigation revealed a rich phenomenology for band touchings in the system. Besides Dirac point band touchings, there are semi-Dirac points (band touchings with a linear dispersion in one direction and quadratic in the other), or higher-energy dispersion points. Among the latter, we show that there is a unique supermerging band touching at $\pm\mb K$ that can be understood from a collision scenario of all possible Dirac points under a variation of the hopping integrals. Their uniqueness in hopping integral parameter space indicates that they are extremely unstable. Moreover, the peculiarity of this point resides in the fact that is characterized by a high-energy dispersion, but a low topological charge. This is due to the fact that the supermerging points result from a union of Dirac points organized in triplets with alternating chirality. Numerical and analytical investigations also revealed a new phenomena in N4 graphene: the formation of Fermi lines for a particular choice of parameters. The particular constraints to obtain them indicate again that they are unstable band touchings.

The creation of multiple Dirac points is a precondition to achieve phases with a large Chern number. This is put to test by implementing the Haldane model in the distant-neighbor hopping graphene. The Haldane mass term gaps the Dirac points such that new QAH phases appear. We have presented various  Chern number phase diagrams to illustrate the role of distant hoppings in the Haldane mass term. The flux dependence allows one to resolve neighbor Dirac points with the different chirality by gapping them with an opposite mass. Said differently, the mass term now changes sign not only between $\mb K$ and $-\mb K$, but also between the satellite created near the regular Dirac points. In principle, for $2n$ Dirac points in the modified graphene, phases with Chern number $|\mc C|=n$ can be created.

As a final remark concerning these N$n$ graphene-Haldane models, 
 we stress that we do not claim that such long-range hopping is relevant to graphene physics.
We believe, however, that  the phenomenology of complex band touchings and large Chern number phases that appears in this  two-band long-range hopping model is rather universal 
and might appear as the effective low energy physics of a more realistic nearest-neigbor model with N orbitals or N atoms per unit cell.
In support of this view there is a recent work\cite{Montambaux2012} that establishes  a mapping of the low  energy physics
of the bilayer graphene (four atoms per unit cell) with that of N3 graphene near supermerging.

\begin{acknowledgments}
The authors would like to thank J.-N. Fuchs, P.\ Simon, C.\ Bena, P.\ Kalugin and G.\ Montambaux  for stimulating discussion.
\end{acknowledgments}
\bibliographystyle{apsrev4-1}
\bibliography{bibl}
\end{document}